\documentclass[twocolumn,amsmath,amssymb,aps,prb,longbibliography]{revtex4-1}
\usepackage{color,graphicx}
\usepackage{amssymb}
\usepackage{float}
\begin{document}

\title{Skyrmion-Skyrmionium Phase Separation and Laning Transitions via Spin-Orbit Torque Currents}
\author{N. P. Vizarim$^1$, J. C. Bellizotti Souza$^2$, C. J. O. Reichhardt$^3$, C. Reichhardt$^3$, P. A. Venegas$^4$, F. B\'eron$^1$}
\affiliation{$^1$ Instituto de F\'isica ``Gleb Wataghin,'' Departamento de F\'isica da Mat\'eria Condensada, Universidade Estadual de Campinas - UNICAMP, 13083-859, Campinas, S\~ao Paulo, Brazil \\
$^2$ POSMAT - Programa de P\'os-Gradua\c{c}\~ao em Ci\^encia e Tecnologia de Materiais, Faculdade de Ci\^encias, Universidade Estadual Paulista - UNESP, Bauru, SP, CP 473, 17033-360, Brazil \\
$^3$ Theoretical Division and Center for Nonlinear Studies, Los Alamos National Laboratory, Los Alamos, New Mexico 87545, USA\\
$^4$ Departamento de F\'isica, Faculdade de Ci\^encias, Universidade Estadual Paulista - UNESP, CP 473, 17033-360 Bauru, SP, Brazil }

\date{\today}

\begin{abstract}
Many driven binary systems can exhibit laning transitions 
when the two species have different mobilities,
such as colloidal particles with opposite charges in electric fields.
Another example is pedestrian or active matter systems,
where particles moving in opposite directions form a phase-separated state that enhances the overall mobility.
In this work, we use atomistic simulations to demonstrate that mixtures of skyrmions and skyrmioniums
also exhibit pattern formation and laning transitions.
Skyrmions move more slowly and at a finite Hall angle compared to skyrmioniums,
which move faster and without a Hall effect.
At low drives, the system forms a partially jammed phase where the skyrmionium is dragged by the surrounding skyrmions,
resulting in a finite angle of motion for the skyrmionium.
At higher drives, the system transitions into a laned state,
but unlike colloidal systems, the lanes in the skyrmion–skyrmionium
mixture are tilted
relative to the driving direction due to the intrinsic skyrmion Hall angle.
In the laned state,
the skyrmionium angle of motion is reversed
when it aligns with the tilted lane structure.
At even higher drives, the skyrmioniums collapse into skyrmions.
Below a critical skyrmion density, both textures can move
independently with few collisions, but
above this density, the laning state disappears entirely,
and the system transitions to a skyrmion-only state.
We map out the velocity and Hall responses of the different textures and identify three distinct phases: partially jammed, laned, and skyrmion-only
moving crystal states.
We compare 
our results to recent observations of tilted laning phases in pedestrian flows,
where chiral symmetry breaking in the particle interactions leads to similar behavior.
\end{abstract}

\maketitle

\vskip 2pc

\section{Introduction}

Recently, there has been great interest in understanding the dynamics
of interacting particles that are not all moving in the same direction.
Assemblies of such particles can undergo structural transitions
when subjected to external driving forces.
In many cases, the particles share the same size and interaction forces;
however, when the particles have
different sizes or interaction strengths,
order-disorder transitions may occur even
in the absence of external driving 
\cite{sadr-lahijany_dispersity-driven_1997,hamanaka_transitions_2006,reichhardt_disordering_2008}.
Moreover, in systems where the particles are of uniform size
but exhibit different dynamical behaviors, disordering transitions and other dynamical phases can emerge.
For instance, in studies of oppositely driven repulsive particles,
the system may form
static ordered states or exhibit fluctuating disordered flow.
At higher driving forces,
these flows may transition into lane forming states
consisting of multiple,
partially phase-separated stripes of oppositely moving particles
\cite{schmittmann_driven_1998,reichhardt_disordering_2019,helbing_freezing_2000,dzubiella_lane_2002,netz_conduction_2003,ikeda_instabilities_2012,wachtler_lane_2016,poncet_universal_2017,bain_critical_2017,reichhardt_laning_2018,bacik_lane_2023}.
The lane arrangement minimizes collisions
and enhances the efficiency of the motion \cite{bacik_lane_2023}.
The observation of lane formation across such diverse systems suggests
that it is triggered by
a universal mechanism.

In general, the lanes align with the direction of driving;
however, Bacik {\it et al.}
recently studied a pedestrian model that included
a chiral symmetry-breaking effect \cite{bacik_lane_2023}.
The system forms tilted lanes in which the flow is at an angle
to the driving direction.
This result was confirmed experimentally by asking actual pedestrians to
deliberately
bias their direction of motion when walking past an oppositely moving
pedestrian, which adds an effective chiral term to the interaction.
Although there have been numerous studies on laning behaviors in
colloidal assemblies, active matter,
dusty plasmas, and other systems,
there are relatively few
studies of the laning of particle-like
magnetic textures such as skyrmions and
skyrmioniums that can be supported within ferromagnets.
If more than one type of magnetic texture is present simultaneously, the
different textures move at different velocities for a given applied
drive,
providing
another type of system that could exhibit laning phenomena.
Unlike colloidal particles, however,
many of the magnetic textures have
a chiral symmetry breaking and move with a Hall angle,
and might thus be expected to form tilted lanes similar to those
observed by 
Bacik {\it et al.} when they added a chiral symmetry
breaking term to their pedestrians.
Additionally, topological transitions among different
magnetic textures can occur if the driving force is sufficiently strong,
so the laning dynamics of magnetic textures could be much richer than
what has been observed in other systems.

Magnetic skyrmions are topologically protected particle-like textures
\cite{muhlbauer_skyrmion_2009,fert_magnetic_2017,bogdanov_thermodynamically_1994,rosler_spontaneous_2006}
that exhibit many similarities to particle-based systems
such as type II superconducting vortices and
charged colloids. Their mutually repulsive interactions
cause the skyrmions to 
form triangular arrays\cite{muhlbauer_skyrmion_2009,yu_real-space_2010}
in order to reduce their interaction energy.
Additionally, skyrmions can be set into motion by external currents \cite{schulz_emergent_2012,jonietz_spin_2010} and can interact with defects or pinning sites,
which inhibit motion below a threshold current \cite{reichhardt_depinning_2016,reichhardt_statics_2022}.
Skyrmions are promising candidates for spintronics
applications, such as memory and novel computing,
due to their
topological protection \cite{bogdanov_thermodynamically_1994,rosler_spontaneous_2006,skyrme_unified_1962},
reduced size
(with diameters ranging from $10$ nm to $\approx1\mu$m \cite{jiang_blowing_2015,yu_real-space_2010,tonomura_real-space_2012}),
and their ability to be transported by spin-polarized currents \cite{schulz_emergent_2012,jonietz_spin_2010}.
Due to their topology \cite{everschor-sitte_real-space_2014},
under external driving skyrmions move at an angle to the applied
drive known as the intrinsic skyrmion Hall angle, $\theta_{\rm sk}^{\rm int}$
\cite{jiang_direct_2017,iwasaki_universal_2013,litzius_skyrmion_2017},
which has a sign that 
depends on the skyrmion winding number, $Q=\pm 1$ \cite{jiang_direct_2017}.
The skyrmion motion
thus has an effective Magnus force component that makes the skyrmion Hall
angle finite and
also affects how
the skyrmion interacts with quenched disorder or other skyrmions. In contrast,
superconducting vortices and colloidal particles typically do not have a
finite Hall angle.


Skyrmionium, another particle-like
magnetic texture, can be viewed as
two skyrmions with opposite polarity and topological charge
that have been combined into a single magnetic texture
\cite{gobel_beyond_2021}. The concept of a skyrmionium first
appeared in the literature as a
``$2\pi$ vortex'' \cite{bogdanov_stability_1999},
and in later works it was referred to
as a ``donut skyrmion'' \cite{streubel_manipulating_2015},
``$2\pi$ skyrmion'' \cite{hagemeister_controlled_2018},
or ``target skyrmion'' \cite{zheng_direct_2017}.
Structurally, a skyrmionium consists of a central skyrmion surrounded 
by an annular domain wall with an opposite topological charge,
resulting in a net topological charge of $Q=0$.
Since the deflection of magnetic textures under an applied drive depends
on the sign and magnitude of $Q$,
skyrmioniums  move without a Hall angle,
making them particularly attractive for applications \cite{kolesnikov_skyrmionium_2018}. 
Skyrmioniums
can be generated in ferromagnetic films using laser pulses
\cite{finazzi_laser-induced_2013,fujita_ultrafast_2017},
and studies have shown that they can move significantly faster than skyrmions
\cite{kolesnikov_skyrmionium_2018,ishida_theoretical_2020,souza_skyrmionium_2025,zhang_control_2016}. 
Methods for transporting skyrmioniums include polarized currents\cite{komineas_skyrmion_2015_1},
magnetic field gradients \cite{komineas_skyrmion_2015},
and spin waves\cite{li_dynamics_2018}. 
Recent studies suggest, however, that skyrmioniums may be less dynamically stable than skyrmions.
Under high driving forces, skyrmioniums can collapse into skyrmions \cite{xia_current-driven_2020,souza_comparing_2024}.
Moreover, in materials with low Gilbert damping,
skyrmioniums are prone to experiencing significant distortions,
further reducing their stability \cite{ishida_theoretical_2020}.

Although numerous studies have addressed the dynamics of skyrmions and skyrmioniums individually,
the behavior of a mixed system containing both skyrmions and skyrmioniums remains unexplored.
The distinct dynamical properties of these textures, such as the presence or absence of a Hall angle,
differences in stability, and variations in size,
could produce a wide variety of order-disorder transitions.
Reichhardt and Reichhardt \cite{reichhardt_laning_2018} studied the dynamics of mixtures of repulsive particles
with varying Magnus terms to model spinning colloidal particles
in fluid flow, which could be applied to
mixtures of skyrmions of different species.
Under an applied external drive in the colloidal system,
there is a critical
driving threshold 
at which a
Magnus-induced disordering transition can occur.
The transition arises because
the differing Hall angles of the two particle species
cause their motions to decouple at the critical drive.
At higher drive values, lane formation
occurs in which the two species separate and move in distinct directions.

In this work, we perform atomistic simulations of skyrmions and skyrmioniums coexisting
within the same ferromagnetic thin film.
For systems with a single skyrmionium among multiple skyrmions, we observe distinct behaviors depending on the skyrmion density. 
At intermediate skyrmion densities, there is an interplay between the motion
of the skyrmions and the skyrmionium.
In this regime, the skyrmionium exhibits a positive
nonzero angle of motion relative to the driving
direction while the skyrmions continue to move
in a direction close to their intrinsic skyrmion Hall angle.
At high skyrmion densities, the skyrmionium is dragged along with the
skyrmions, moving at an angle closer to the intrinsic
skyrmion Hall angle.
Interestingly, for intermediate skyrmion densities, the difference
in velocity between the skyrmions and the skyrmionium
permits the skyrmionium
to open a dynamic lane through the skyrmions.
In this state, the skyrmionium can travel with a high velocity,
and there is a clear phase separation
between the skyrmion and skyrmionium species.
When additional skyrmioniums are introduced into the system,
they can follow and reinforce
skyrmionium lanes that have been established through
the skyrmions, thereby enhancing their collective motion. 
Beyond skyrmion and skyrmionium mixtures,
other possible mixed magnetic textures include
skyrmion–antiskyrmion combinations \cite{zheng_skyrmionantiskyrmion_2022,wang_nonvolatile_2025},
skyrmions and skyrmion bundles \cite{zhang_stable_2024},
skyrmions and chiral bobbers \cite{zheng_experimental_2018},
and coexisting tubular and partial skyrmion phases \cite{mandru_coexistence_2020}.
Similarly, in liquid crystal systems, different textures with
varying mobilities
can coexist
\cite{sohn_schools_2019,coelho_sculpting_2023,amaral_liquid_2025}.
These findings suggest that magnetic textures are a fruitful 
class of systems in which to study lane formation effects.

\section{Simulation}

\begin{figure}
\centering
\includegraphics[width=0.7\columnwidth]{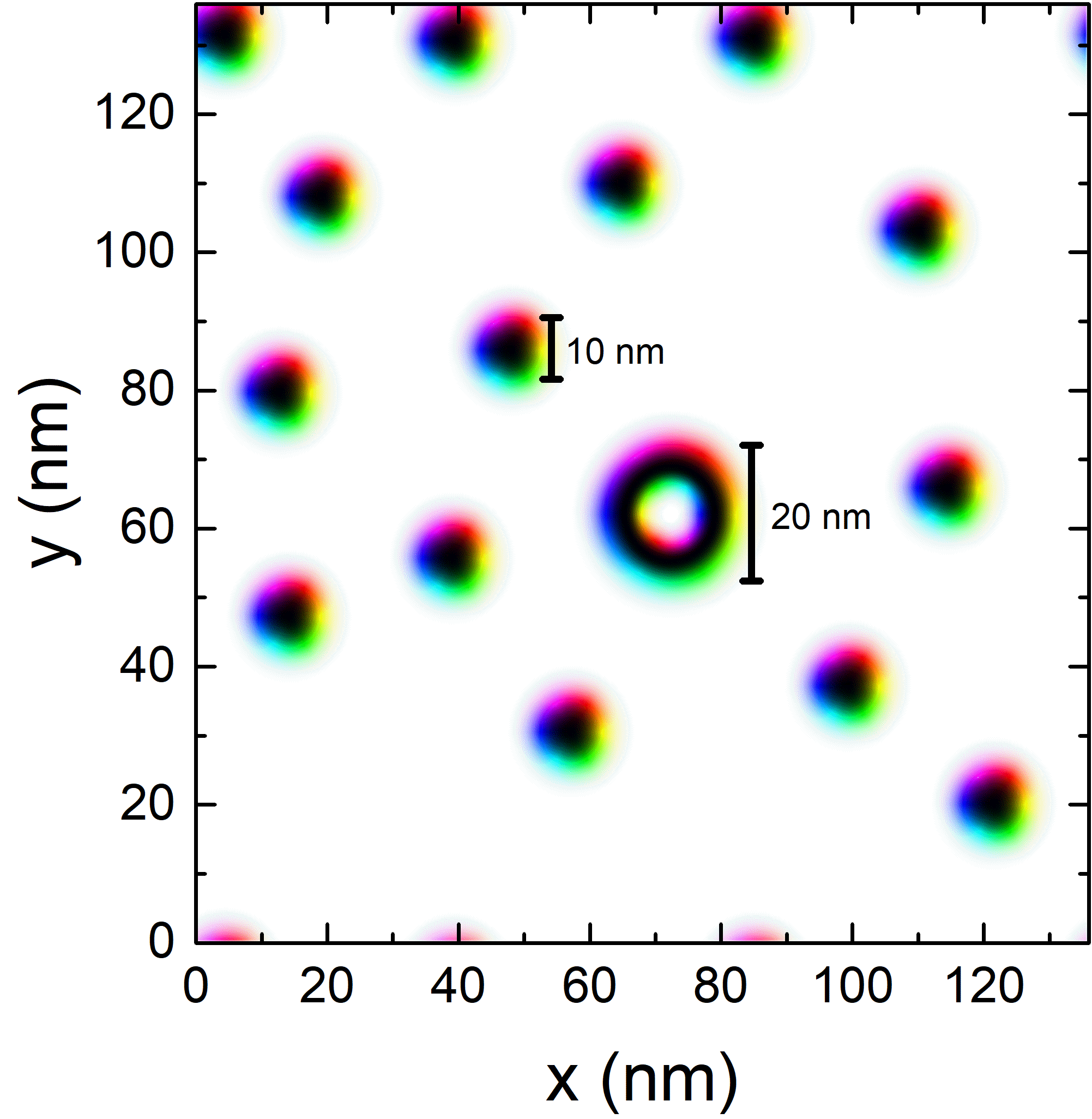}
\caption{
Illustration of a representative sample
containing $N_{\rm sk}=14$ skyrmions and
$N_{\rm skium}=1$ skyrmionium after the stabilization process
at zero current,
$j=0 \text{ A/m}^{2}$.
The magnetic textures are visualized by projecting
the magnetic moments onto the $xy$ plane.
The skyrmions and skyrmionium were initially placed at random positions,
and their average diameters are approximately $10$ and $20$ nm, respectively.
}
\label{Fig1}
\end{figure}

We simulate the collective behavior of
$N_{\rm skium}$ skyrmioniums among $N_{\rm sk}$ skyrmions
in an ultrathin ferromagnetic material
at $T=0$ K
that can host Néel skyrmions. The film has dimensions
of $L \times L$, where $L= 136$ nm,
and has periodic boundary conditions along
the $x$ and $y$ directions.
A magnetic field is applied
perpendicular to the film surface.
The skyrmionium density is $n_{\rm skium}=N_{\rm skium}/L^2$ and 
the skyrmion density is $n_{\rm sk}=N_{\rm sk}/L^2$. 
We perform the simulations for several
ratios of $N_{\rm skium}/N_{\rm sk}$ in order to characterize
the skyrmionium-skyrmion behavior under the application of
a spin-orbit-torque (SOT) current.
In Fig.~\ref{Fig1} we illustrate a representative sample
containing $N_{\rm skium}=1$ skyrmionium and
$N_{\rm sk}=14$ skyrmions.

The skyrmion and skyrmionium behavior is described using the atomistic
model \cite{evans_atomistic_2018}, which captures the dynamics of
individual atomic magnetic moments.
This method allow us to investigate the spin textures in detail.
The Hamiltonian governing the atomistic
dynamics is given by
\cite{evans_atomistic_2018, iwasaki_universal_2013, iwasaki_current-induced_2013}:

\begin{align}\label{eq1}
  \mathcal{H}=&-\sum_{i, j\in n.n.}J_{ij}\mathbf{m}_i\cdot\mathbf{m}_j
                -\sum_{i, j\in n.n.}\mathbf{D}_{ij}\cdot\left(\mathbf{m}_i\times\mathbf{m}_j\right)\\\nonumber
                &-\sum_i\mu\mathbf{H}\cdot\mathbf{m}_i
                -\sum_iK(\mathbf{m}_i\cdot \mathbf{\hat{z}})^2 \ .
\end{align}
The underlying lattice is a square arrangement of atomic spins with lattice
constant $a=0.5$ nm.
The first term on the right side of Eq.~(1) is the exchange interaction
between nearest neighbor magnetic moments $i$ and $j$
with an exchange constant of $J_{ij}=J$.
The second term is the interfacial Dzyaloshinskii–Moriya (DM)
interaction, where $\mathbf{D}_{ij}=D\hat{{\bf z}}\times\hat{\bf{r}}_{ij}$ is the
DM
vector between magnetic moments $i$ and $j$, $D$ is the DM constant,
and $\hat{{\bf r}}_{ij}$ is the unit distance
vector between atomic sites $i$ and $j$.
The DM term is what stabilizes the N\'eel skyrmions
and skyrmioniums.
The third term is the Zeeman interaction with an applied external magnetic
field $\mathbf{H}$, and the fourth term is the sample anisotropy with strength $K$.
Here $\mu=\hbar\gamma$ is the magnitude of the magnetic moment
and $\gamma=1.76\times10^{11}$ T$^{-1}$ s$^{-1}$ is the electron
gyromagnetic ratio.
Since we are considering ultrathin films, long-range dipolar interactions
can be neglected as they are expected to be very small
\cite{paul_role_2020}.

The time evolution for the individual
atomic magnetic moments is obtained from the Landau-Lifshitz-Gilbert equation augmented with
the SOT current \cite{seki_skyrmions_2016, gilbert_phenomenological_2004}:
\begin{equation}\label{eq2}
    \frac{\partial\mathbf{m}_i}{\partial t}=-\gamma\mathbf{m}_i\times\mathbf{H}^\text{eff}_i
                             +\alpha\mathbf{m}_i\times\frac{\partial\mathbf{m}_i}{\partial t}
                             +\frac{j\hbar\gamma Pa^2}{2e\mu}\mathbf{m}\times(\mathbf{\hat{j}}\times\mathbf{\hat{z}})\times\mathbf{m} \ .
\end{equation}
Here 
$\mathbf{H}^\text{eff}_i=-\frac{1}{\hbar\gamma}\frac{\partial \mathcal{H}}{\partial \mathbf{m}_i}$
is the effective magnetic field, including all interactions from
the Hamiltonian, $\alpha$ is the phenomenological damping
introduced by Gilbert, and the last term is the torque induced by
the SOT current, where $j$
is the current density, $P=1$ is the spin polarization,
$e$ is the electron charge,
and $a$ is the lattice parameter.
We fix $\mu\mathbf{H}=0.5(D^2/J)\mathbf{\hat{z}}$ and
$\alpha=0.4$. In all cases, the external current ${\bf j}$ is applied along $-\hat{{\bf x}}$, which results in
a driving force along $-\hat{{\bf x}}$ in both magnetic textures;
however, since
skyrmions have a topological charge
of $Q=\pm 1$, they exhibit a
finite skyrmion Hall angle
\cite{zhang_control_2016, jiang_direct_2017, souza_comparing_2024}.
The material parameters
are $J=1$ meV,
$D=0.2J$, and $K=0.01J$. These material parameters
stabilize Néel skyrmions similar to those found for Pt/Co/MgO thin-films
\cite{boulle_room-temperature_2016}.

For each simulation, we place
a selected number of skyrmions and skyrmioniums randomly throughout the sample
and perform
a simulated annealing process
using a stochastic gradient descent (SGD) method
\cite{kiefer_stochastic_1952,robbins_stochastic_1951}.
The system settles into a ground state that we use as input for the dynamics.
The numerical integration of Eq.~\ref{eq2} is performed using
a fourth order Runge-Kutta method.
For each value of $j$,
we calculate the time average velocity
components, $\left\langle v_x\right\rangle_{\rm sk,skium}$ 
and $\left\langle v_y\right\rangle_{\rm sk,skium}$,
for the skyrmions and skyrmionium(s), respectively,
over $200$ ns to ensure a steady state measurement.

\section{Single skyrmionium among multiple skyrmions}

\begin{figure}
\centering
\includegraphics[width=0.8\columnwidth]{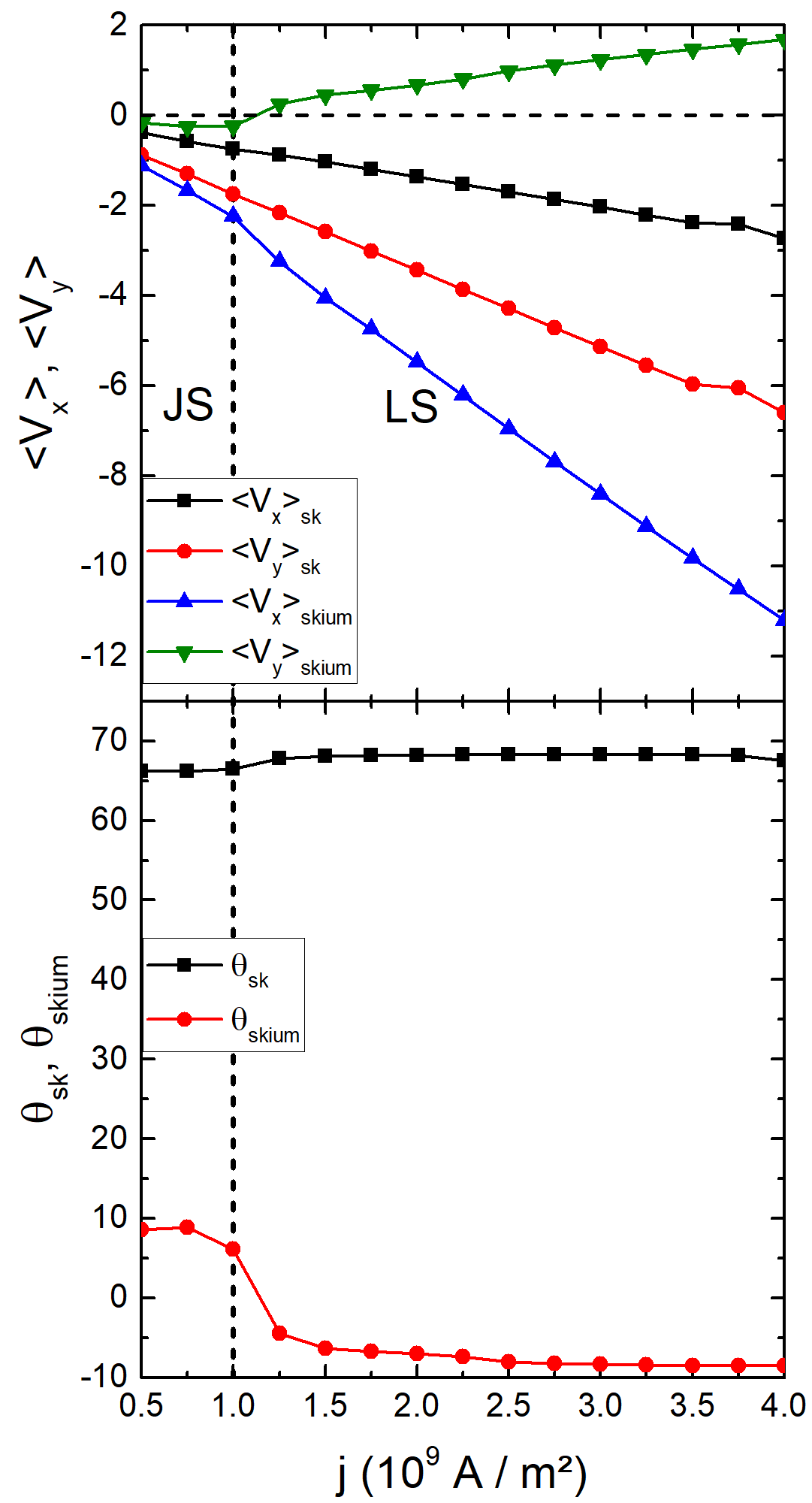}
\caption{
(a) The velocity components $\langle V_x\rangle_{\rm skium,sk}$ and $\langle V_y\rangle_{\rm skium,sk}$
and (b) the corresponding angle of motion
$\theta_{\rm skium,sk}$
with respect to the drive
vs 
the external drive $j$ for the
sample illustrated in Fig.~\ref{Fig1} with
$N_{\rm skium}=1$ skyrmionium among
$N_{\rm sk}=14$ skyrmions.
The dashed vertical line separates the jammed state (JS)
from the laning state (LS).
The dashed horizontal line in (a)
highlights the transition from negative to positive velocity
values.
}
\label{Fig2}
\end{figure}

\begin{figure}
 \centering
 \includegraphics[width=1.0\columnwidth]{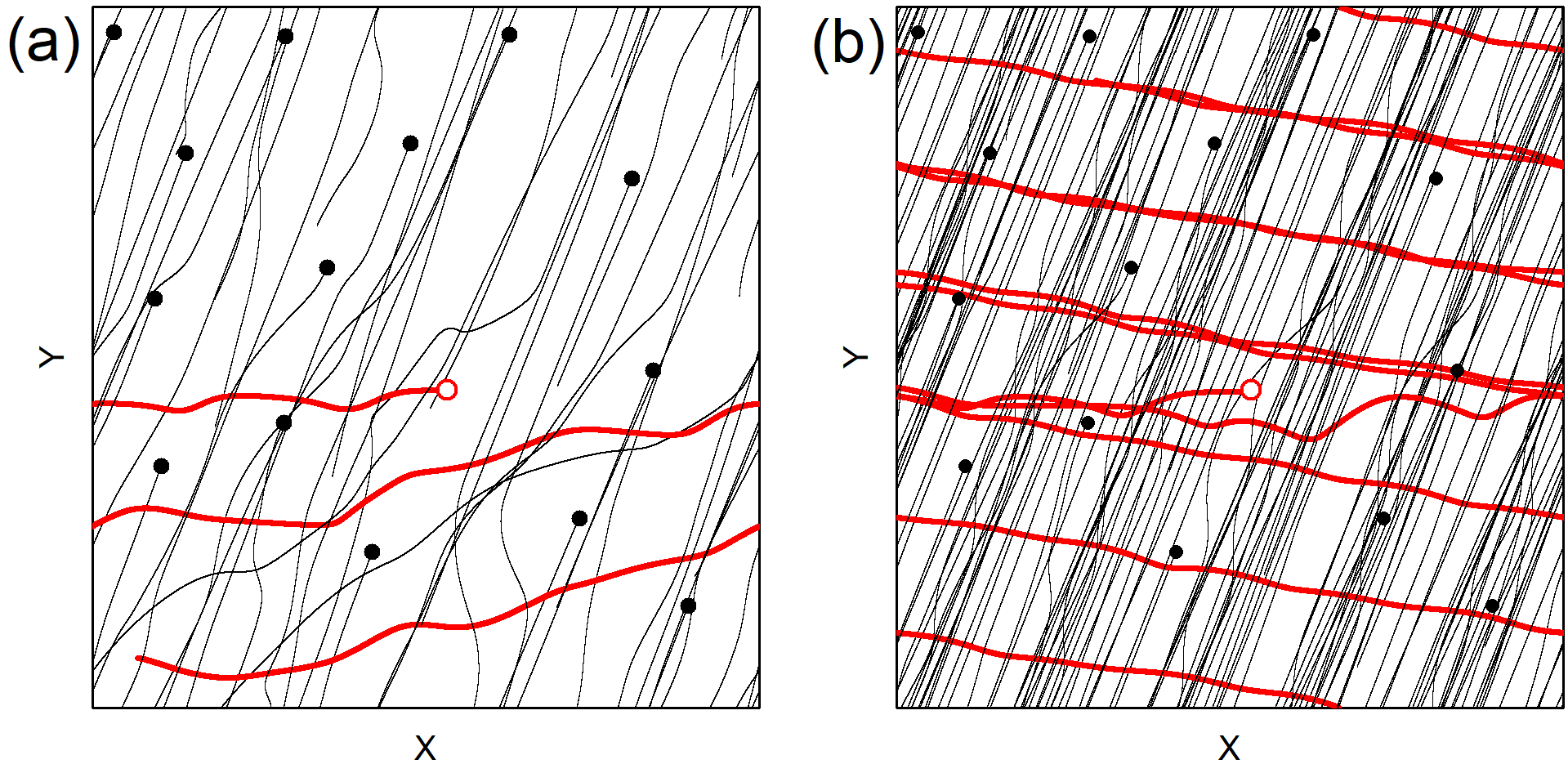}
 \caption{
 Illustration of the skyrmion (black) and skyrmionium (red) trajectories for the sample from
 Fig.~\ref{Fig1} with $N_{\rm skium}=1$ and $N_{\rm sk}=14$ over a
 time interval of
 $\Delta t =200$ ns. 
 (a) The JS at $j=0.75 \times 10^{9} \text{ A/m}^{2}$.
 (b) The LS at $j=1.50 \times 10^{9} \text{ A/m}^{2}$.
 The large black dots indicate the initial positions of the skyrmions,
 and the open red circle is the initial position of
 the skyrmionium.
 }
   \label{Fig3}
\end{figure}

We first consider a single skyrmionium
stabilized among 14 skyrmions.
Figure~\ref{Fig1} 
shows the static configuration of this system under
zero external drive, $j=0 \text{ A/m}^{2}$.
In Fig.~\ref{Fig2}
we plot the average skyrmionium and skyrmion velocity components 
$\left\langle V_x\right\rangle_{\rm sk,skium}$ 
and $\left\langle V_y\right\rangle_{\rm sk,skium}$
along with
the skyrmionium and skyrmion angle of motion
$\theta_{\rm sk,skium}$
with respect to
the drive
as a function of the external current, $j$.
We find that
the skyrmion velocity components
$\langle V_x\rangle_{\rm sk}$ and
$\langle V_y\rangle_{\rm sk}$
increase monotonically with drive.
In contrast,
for the skyrmionium,
$\langle V_y\rangle_{\rm skium}$ exhibits a sudden reversal in sign at 
$j=1.25\times 10^{9} \text{ A/m}^{2}$, indicating
that the skyrmionium has changed its direction of motion.
At this same current,
$\langle V_x\rangle_{\rm skium}$
increases in magnitude.
Figure~\ref{Fig2}(b)
shows that the direction of motion of the skyrmions barely changes
with drive but remains very close to
$\theta_{\rm sk} \approx 67^{\circ}$ for all values of $j$.
The skyrmionium direction
of motion, on the other hand,
exhibits a complete reversal from
$\theta_{\rm skium} \approx 9^{\circ}$ to 
$\theta_{\rm skium} \approx -9^{\circ}$. 
The velocity and direction of motion changes
exhibited by the skyrmionium are signatures of 
a dynamic transition.
For $j \leq 1.00\times 10^{9} \text{ A/m}^{2}$ the system
is in a partially jammed state (JS)
where the skyrmionium is dragged along by
the skyrmions, giving it a finite and positive
angle of motion.
When the skyrmionium collides with
the skyrmions, it deviates from its natural trajectory, 
which in the absence of other textures or defects would be along
$\theta_{\rm skium} = 0^{\circ}$ \cite{kolesnikov_skyrmionium_2018}.
The jamming effect is relatively weak due to the moderate skyrmion density.
For $j>1.00\times 10^{9} \text{ A/m}^{2}$, the system transitions
to a laning state (LS), where the skyrmionium
is able to open a path
through the skyrmions and flow at higher velocity while experiencing
few to no collisions with the skyrmions.
In Fig.~\ref{Fig3}(a) we illustrate
the skyrmionium and skyrmion trajectories for the
partially jammed state at
$j=0.75 \times 10^{9} \text{ A/m}^{2}$ and
in Fig.~\ref{Fig3}(b) we show the same for the
laning state at $j=1.50 \times 10^{9} \text{ A/m}^{2}$.
In the partially jammed state,
the skyrmionium moves along the $-x$ and $-y$ directions, and its motion
is tortuous due to the collisions with the skyrmions that
are traveling along
$-x$ and $-y$ at a larger angle of motion.
The skyrmion trajectories are also tortuous due to collisions with the 
skyrmionium and other skyrmions.
In the laning state,
the skyrmionium moves
along the $-x$ and $+y$ direction,
as also shown in Fig.~\ref{Fig2}.
The trajectories of
both the skyrmionium and skyrmions are less tortuous
since the opening of a lane for skyrmionium motion minimizes the
number of skyrmionium-skyrmion collisions that occur.

\begin{figure}
\centering
\includegraphics[width=1.0\columnwidth]{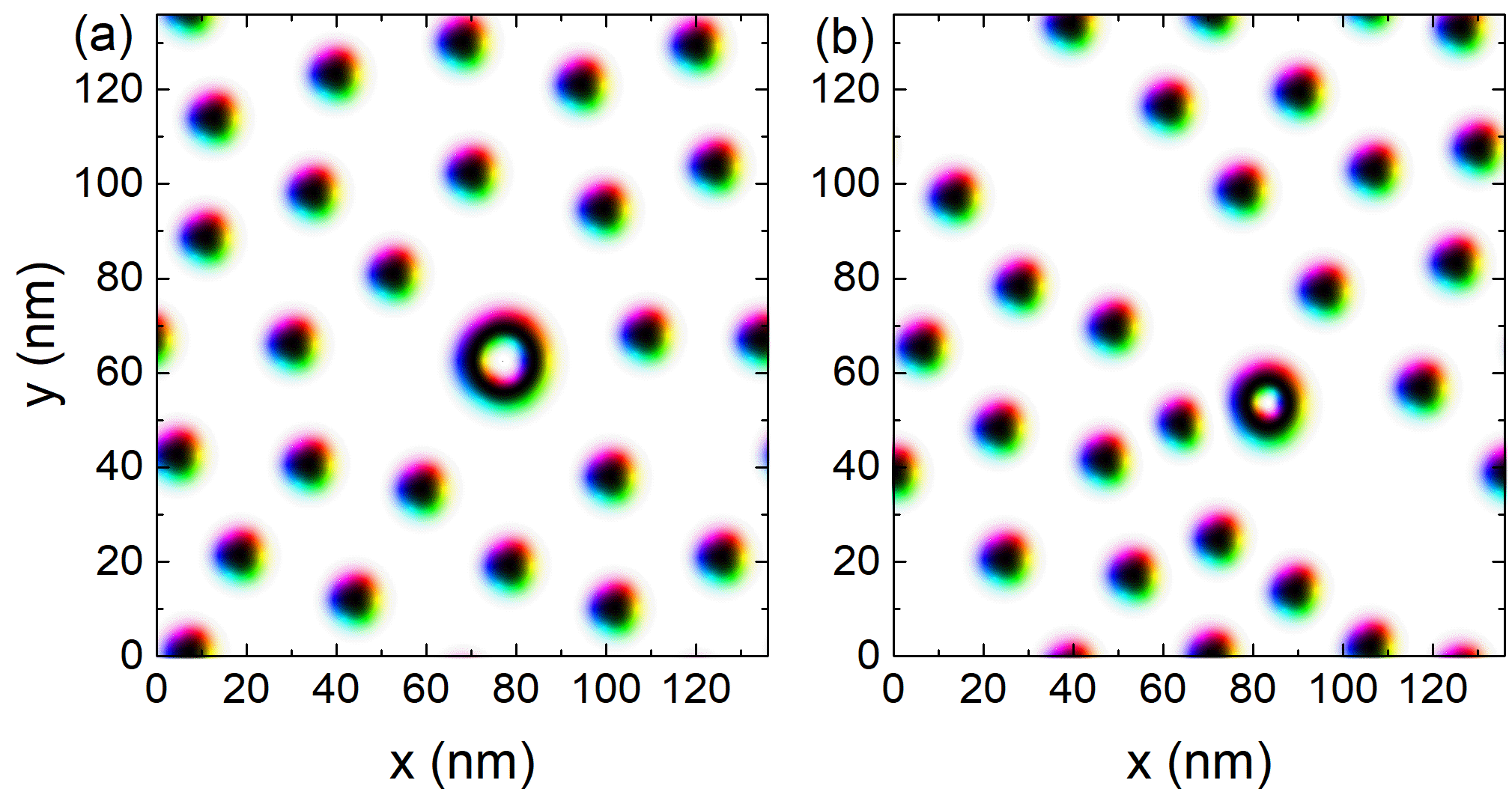}
\caption{
Snapshots of a system with $N_{\rm sk}=25$ and $N_{\rm skium}=1$.
(a) The ground state at $j=0 \text{ A/m}^{2}$.
(b) The laning state at $j=2.25\times 10^{9} \text{ A/m}^{2}$.
The magnetic textures are visualized by projecting
the magnetic moments onto the $xy$ plane.
}
\label{Fig4}
\end{figure}

In Fig.~\ref{Fig4}(a) we show the $j=0$ ground state of
a system with $N_{\rm sk}=25$ and $N_{\rm skium}=1$.
Although the number of skyrmions in the sample is now larger,
the skyrmionium and skyrmion sizes remain 
roughly the same as
in the $N_{\rm sk}=14$ and $N_{\rm skium}=1$ sample
shown in Fig. \ref{Fig1}.
Note that the empty space surrounding the skyrmionium is relatively
larger than that separating adjacent skyrmions.
This is the result of the
larger repulsive interaction exhibited by
skyrmionium \cite{souza_comparing_2024}.

\begin{figure}
\centering
\includegraphics[width=0.8\columnwidth]{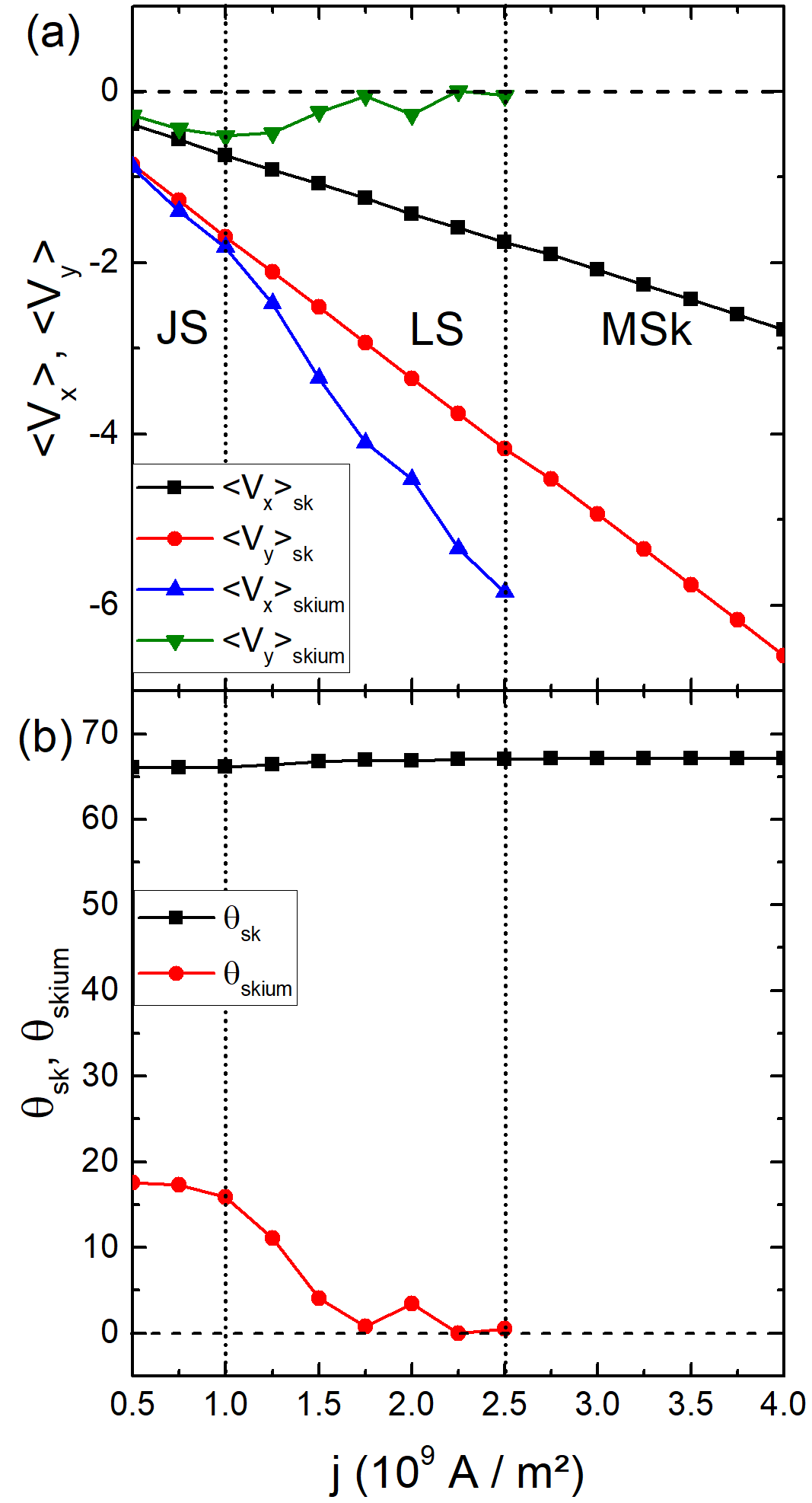}
\caption{
(a) The velocity components $\langle V_x\rangle_{\rm skium,sk}$
and $\langle V_y\rangle_{\rm skium,sk}$
and (b) the corresponding angle of motion $\theta_{\rm skium,sk}$
with respect to the external drive $j$ plotted vs
$j$ for the $N_{\rm skium}=1$ and $N_{\rm sk}=25$ sample from
Fig.~\ref{Fig4}(a).
The dashed vertical lines separate the
partially jammed state (JS), laning state (LS), 
and moving skyrmion lattice (MSk) state.
The dashed horizontal line in (a) highlights the transition from negative
to positive velocity values.
}
\label{Fig5}
\end{figure}

In Fig.~\ref{Fig5} we plot $\langle V_x\rangle_{\rm sk,skium}$,
$\langle V_y\rangle_{\rm sk,skium}$ and $\theta_{\rm sk,skium}$ versus $j$
for the $N_{\rm sk}=25$ and $N_{\rm skium}=1$ system.
Figure~\ref{Fig5}(a) shows that
the skyrmion velocity components $\langle V_x\rangle_{\rm sk}$ and
$\langle V_y\rangle_{\rm sk}$ increase linearly in magnitude with
increasing drive, very similar to the behavior that appears in
Fig.~\ref{Fig2}. 
Here the skyrmions dominate the
behavior of the sample, and their dynamics remain largely unaffected 
by the presence of the skyrmionium.
Similarly, 
in Fig.~\ref{Fig5}(b), the skyrmion
angle of motion $\theta_{\rm sk} \approx 67^{\circ}$ for all values of $j$. 
As more skyrmions are added and the skyrmion density increases,
the influence of the skyrmionium on the skyrmion dynamics
diminishes even further.
We find that
for the skyrmionium, $\langle V_x\rangle_{\rm skium}$
and $\langle V_y\rangle_{\rm skium}$ are separable
into two regimes: a partially jammed state for 
$j < 1.25 \times 10^{9} \text{ A/m}^{2}$, and
a laning state for $j \geq 1.25 \times 10^{9} \text{ A/m}^{2}$.
As was the case in Fig.~\ref{Fig2},
in the partially jammed state the skyrmionium is dragged along by
the skyrmions and moves along a nonzero angle to the driving direction.
Due to the higher skyrmion density in the $N_{\rm sk}=25$ sample
compared to the $N_{\rm sk}=14$ sample,
$\theta_{\rm skium} \approx 18^{\circ}$ in the
partially jammed state
of Fig.~\ref{Fig5}(b) is
larger than the value
$\theta_{\rm skium} \approx 9^{\circ}$ found in Fig.~\ref{Fig2}(b).
In addition, in the laning state of Fig.~\ref{Fig5}(a),
due to the higher skyrmion density
$\langle V_y\rangle_{\rm skium}$ approaches zero but never becomes
positive.
As the external drive $j$ increases in magnitude,
the skyrmions and skyrmionium both travel at higher velocities,
causing stronger deformations and shrinkages
to occur during each collision.
The consequences are illustrated
in Fig.~\ref{Fig4}(b)
at $j=2.25 \times 10^{9} \text{ A/m}^{2}$,
where the laning state is well-established, but where
collisions with skyrmions have caused
the skyrmionium to shrink from its
ground state size of $d \approx 20$ nm to only $d \approx 15$ nm.
The skyrmionium moves at a much higher velocity than the skyrmions,
which facilitates the phase separation,
but also increases the pressure on the skyrmionium
due to the nearly constant contact with the
skyrmions at the edge of the lane.
For $j > 2.50 \times 10^{9} \text{ A/m}^{2}$, the pressure on the skyrmionium
becomes so large
that the internal skyrmion of the skyrmionium annihilates, transforming
the skyrmionium into an ordinary skyrmion.
This is why no skyrmionium velocity or angle of motion measurements
appear in Fig.~\ref{Fig5} above
$j > 2.50 \times 10^{9} \text{ A/m}^{2}$.
The resulting state,
where only coherently moving skyrmions are present in the system,
is called the moving skyrmion lattice (MSk).

\begin{figure}
\centering
\includegraphics[width=1.0\columnwidth]{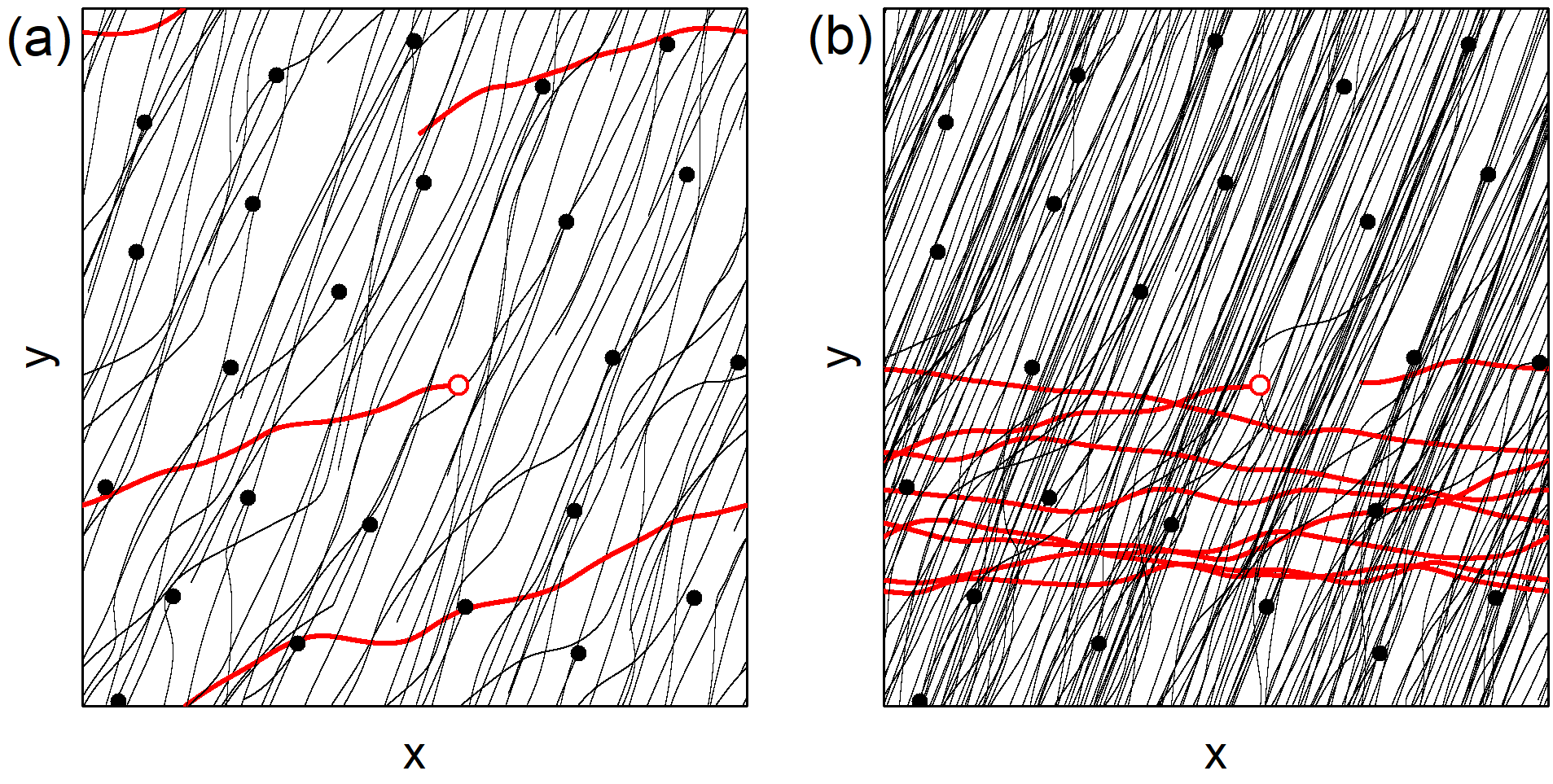}
\caption{
  Illustration of the skyrmion (black) and skyrmionium (red) trajectories for the sample from Fig.~\ref{Fig4} with $N_{\rm skium}=1$ and
  $N_{\rm sk}=25$ over a time interval of
$\Delta t =200$ ns. 
  (a) The JS at $j=0.75 \times 10^{9} \text{ A/m}^{2}$.
  (b) The LS at $j=2.25 \times 10^{9} \text{ A/m}^{2}$.
  The large black dots indicate the initial positions of the skyrmions,
  and the open red circle is the initial position of the skyrmionium.
}
 \label{Fig6}
\end{figure}

For the $N_{\rm skium}=1$ and $N_{\rm sk}=25$ sample,
the skyrmion trajectories are illustrated in
the JS at $j= 0.75 \times 10^{9} \text{ A/m}^{2}$ in
Fig.~\ref{Fig6}(a)
and in 
the LS for $j = 2.25 \times 10^{9} \text{ A/m}^{2}$ in
Fig.~\ref{Fig6}(b).
The behavior is similar to that found in
Fig.~\ref{Fig3}.
In the JS shown in Fig.~\ref{Fig6}(a),
the trajectories are more tortuous due to
the skyrmion-skyrmion collisions and the dragging of the
skyrmionium by the skyrmions.
In the LS shown in Fig.~\ref{Fig6}(b),
the trajectories are straighter and the skyrmionium moves with an
average angle of
$\theta_{\rm skium} \approx 0^{\circ}$.

\begin{figure}
\centering
\includegraphics[width=0.7\columnwidth]{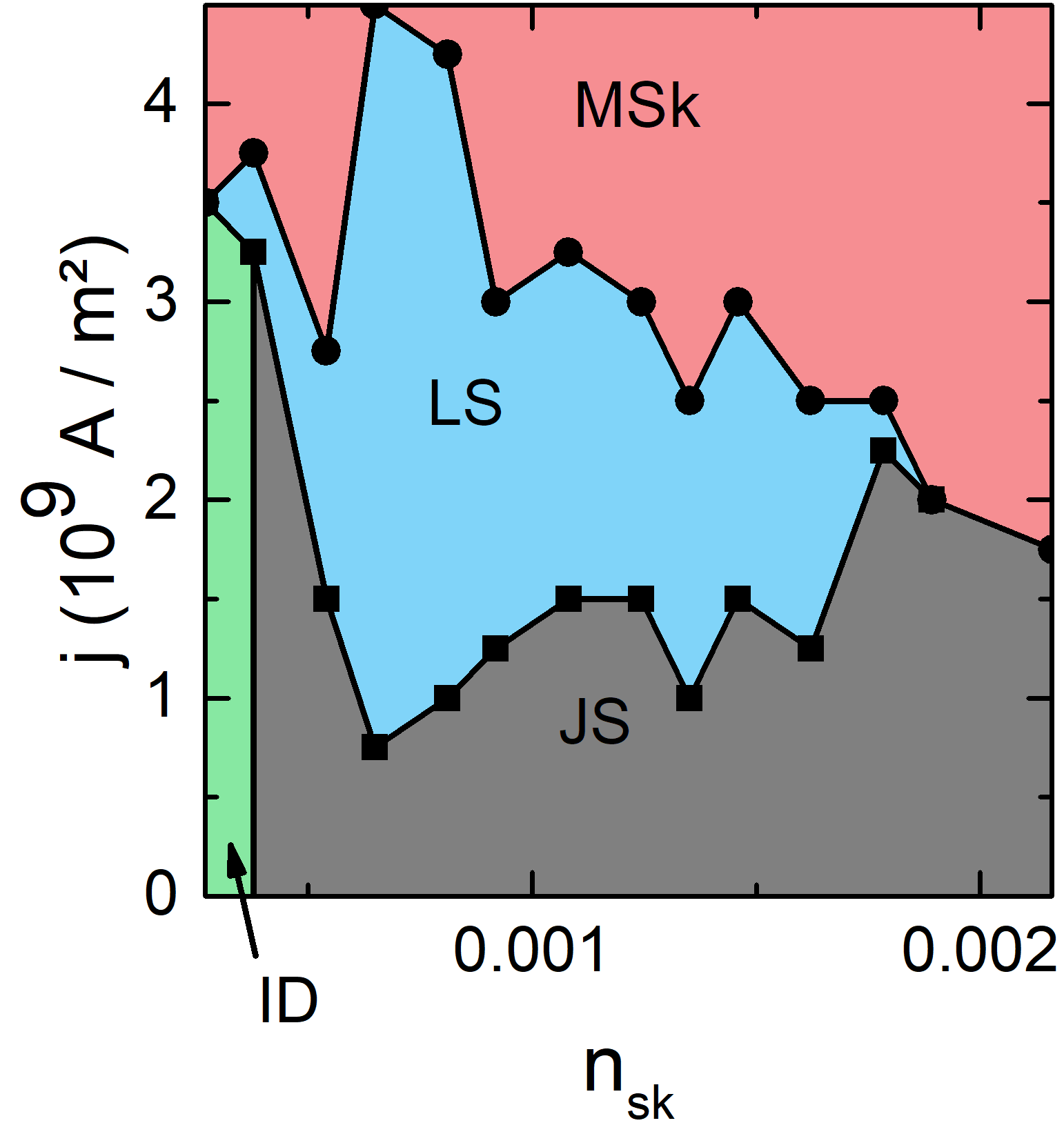}
\caption{
Dynamic phase diagram as a function of $j$ vs skyrmion density $n_{\rm sk}$ for
samples containing $N_{\rm skium}=1$ skyrmionium.
Green: independent dynamics (ID);
gray: partially jammed state (JS);
light blue: laning state (LS);
red: moving skyrmion lattice (MSk).
}
\label{Fig7}
\end{figure}

We performed a series of
simulations for varied currents $j$ and skyrmion densities $n_{\rm sk}$
in order to construct the dynamic phase diagram shown
in Fig.~\ref{Fig7}.
The LS appears in the range
$0.00027 < n_{\rm sk} < 0.00189$.
If the skyrmionium is placed in a very sparse system with a
low density of skyrmions,
both textures move independently
and undergo very few collisions, a regime that
we label as independent dynamics (ID).
This phase
is stable for very low skyrmion densities, $n_{\rm sk} \leq 0.00038$.
Lane formation is still possible in the ID state
because the skyrmionium can influence the motion of the
skyrmions through collisions and generate a phase separation between
the textures. The direction of motion of the skyrmionium is,
however, barely changed in the ID regime, so the system is not
behaving as a partially jammed solid.
When the skyrmion density is large,
$n_{\rm sk} \geq 0.00189$,
the skyrmionium is no longer able to separate itself
from the skyrmions through a laning process
and is strongly dragged by the skyrmion lattice
in the JS 
until skyrmionium annihilation occurs for higher drives.
Phase separation can only occur when the driving is sufficiently large.
For $j < 0.75 \times 10^{9} \text{ A/m}^{2}$,
the LS never appears since the skyrmionium does
not have sufficient energy to
separate from the skyrmion phase at any skyrmion density.
At $n_{\rm sk}=0.00065$, lane formation is optimized and the LS
extends from
$j = 0.75 \times 10^{9} \text{ A/m}^{2}$
all the way up to
$j = 4.50 \times 10^{9} \text{ A/m}^{2}$.
This is very similar to the lane formation
discussed in detail in Figs.~\ref{Fig2} and \ref{Fig3}
for $n_{\rm sk}=0.00081$.
If the external current becomes too strong, however, the
internal skyrmion of the skyrmionium is annihilated
due to collisions with skyrmions,
causing the skyrmionium to transform into a skyrmion and resulting in
the appearance of a
moving skyrmion lattice phase (MSk).

\section{Multiple skyrmioniums among multiple skyrmions}

\begin{figure}
 \centering
 \includegraphics[width=0.8\columnwidth]{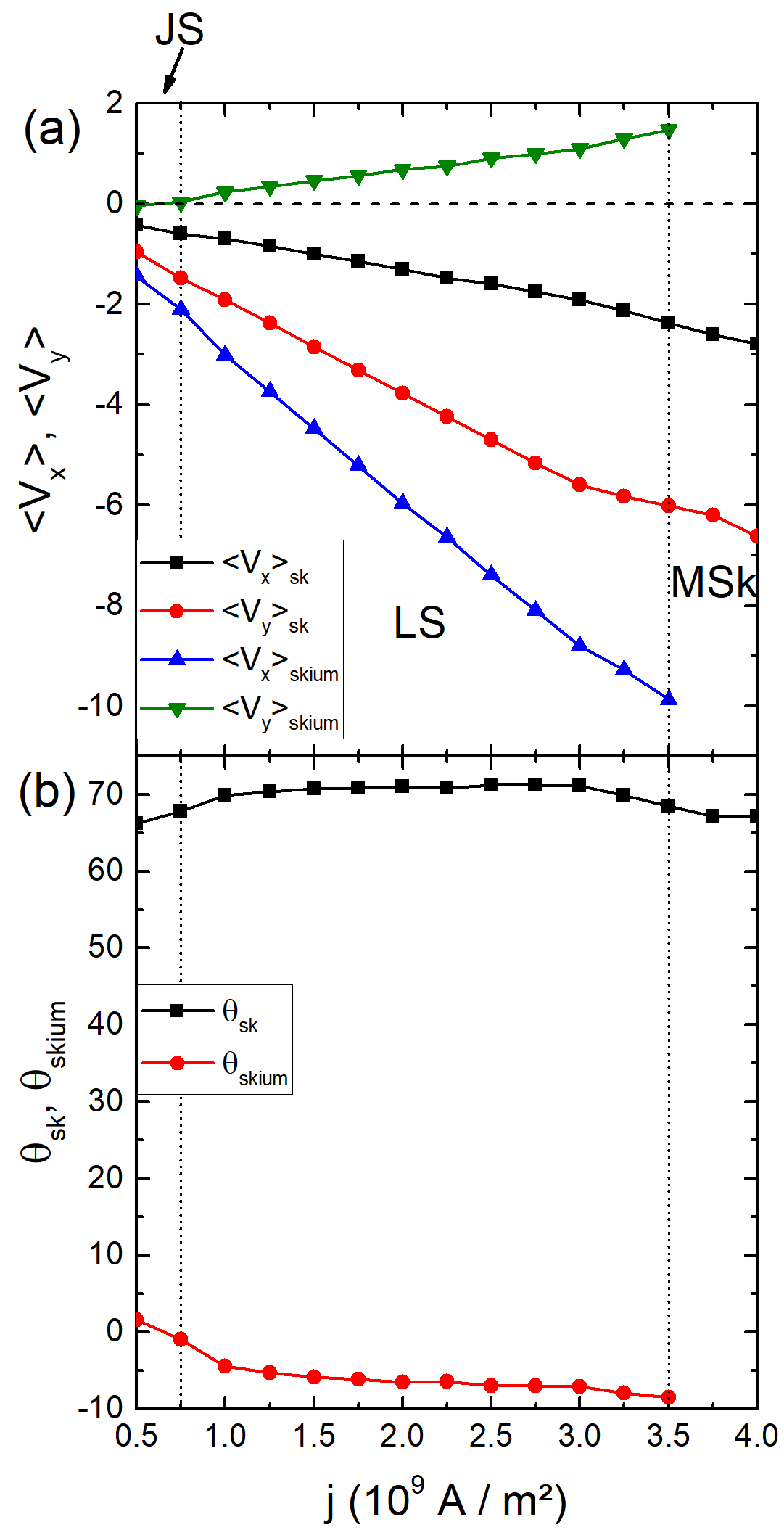}
 \caption{
  (a) The velocity components $\langle V_x\rangle_{\rm skium,sk}$ and $\langle V_y\rangle_{\rm skium,sk}$
   and (b) the corresponding angle of motion
   $\theta_{\rm skium,sk}$ with respect to the external drive $j$
   plotted vs $j$ for the $N_{\rm skium}=3$ and $N_{\rm sk}=10$ sample
   illustrated in Fig.~\ref{Fig9}(a).
   The dashed vertical lines separate the jammed state (JS), laning state (LS) 
    and moving skyrmion lattice (MSk).
 The dashed horizontal line in (a) highlights the transition from negative to positive velocity
    values.
 }
   \label{Fig8}
\end{figure}

\begin{figure}
 \centering
 \includegraphics[width=1.0\columnwidth]{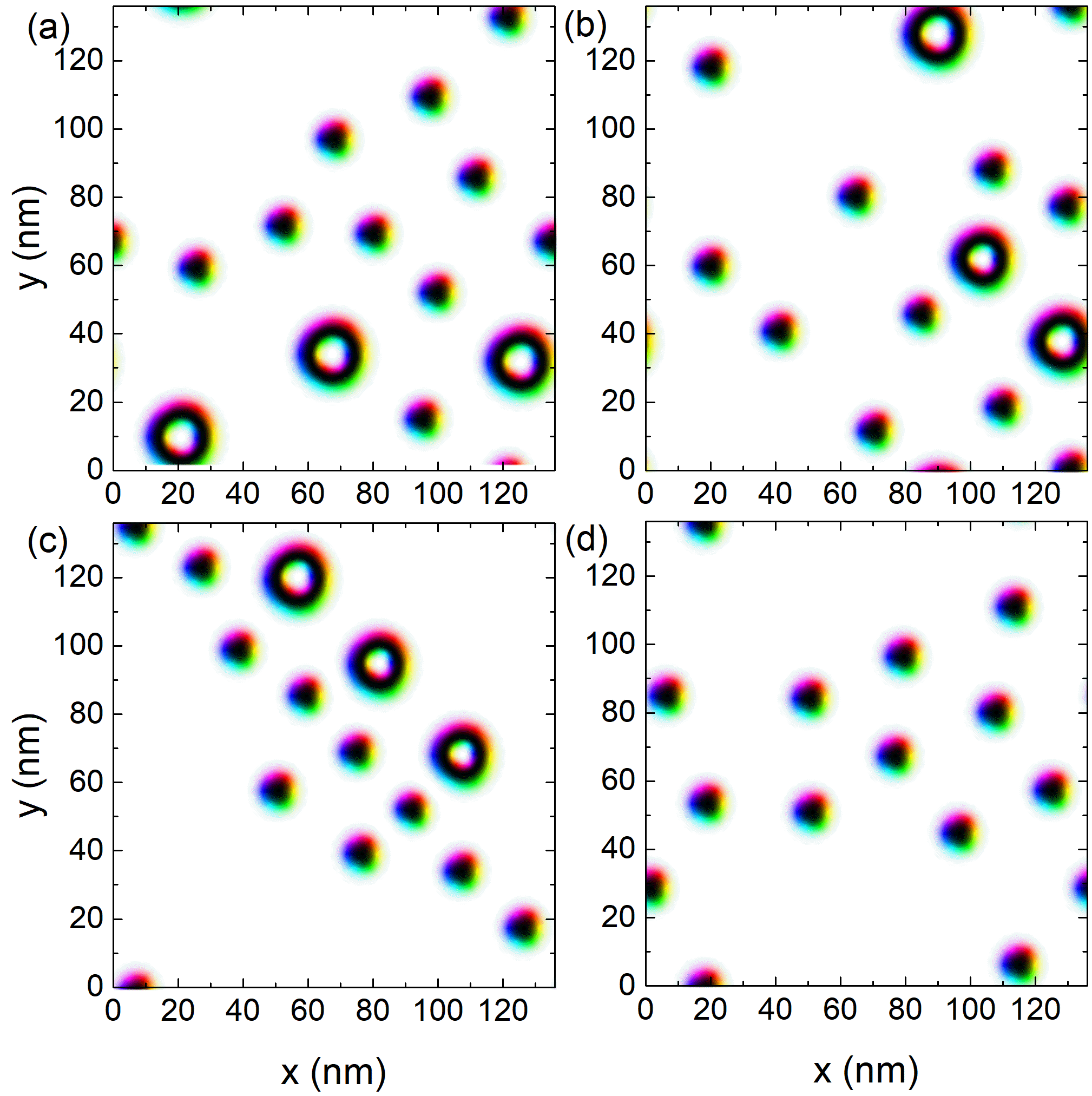}
 \caption{
Snapshots of a system with $N_{\rm sk}=10$ and $N_{\rm skium}=3$.
(a) The ground state at $j=0 \text{ A/m}^{2}$.
(b) The partially jammed state (JS) at $j = 0.75\times 10^{9} \text{ A/m}^{2}$.
(c) The laning state (LS) at $j = 2.25\times 10^{9} \text{ A/m}^{2}$.
(d) The moving skyrmion lattice state (MSk) at $j = 3.75 \times 10^{9} \text{ A/m}^{2}$.
The magnetic textures are visualized by projecting the magnetic moments onto the $xy$ plane.
 }
 \label{Fig9}
\end{figure}

We now move from the single skyrmionium regime to multiple
skyrmioniums.
In Fig.~\ref{Fig8} we plot
$\left\langle V_x\right\rangle_{\rm sk,skium}$,
$\left\langle V_y\right\rangle_{\rm sk,skium}$, and $\theta_{\rm sk,skium}$
versus $j$ for a system with $N_{\rm skium}=3$ and $N_{\rm sk}=10$.
The skyrmion velocity components $\left\langle V_x\right\rangle_{\rm sk}$ and 
$\left\langle V_y\right\rangle_{\rm sk}$
in Fig.~\ref{Fig8}(a) are no longer linear because the dynamics
of the skyrmions are now strongly impacted by the presence of the
skyrmioniums.
The skyrmion angle of motion is also affected,
as illustrated in the plot of $\theta_{\rm sk}$
versus $j$ in Fig.~\ref{Fig8}(b),
where the skyrmion angle oscillates around
$\theta_{\rm sk} \approx 69^{\circ}$.
The laning state (LS) appears at low drives,
$j > 0.75 \times 10^{9}\text{ A/m}^{2}$.
As more skyrmioniums are added to the sample,
it becomes easier for them to establish stable
phase separated lanes passing
through the skyrmions since each skyrmionium can reinforce the lane opened
by the skyrmionium traveling ahead of it.
We show
snapshots from the system in Fig.~\ref{Fig8} at
zero external drive in Fig.~\ref{Fig9}(a),
in the JS where the skyrmions and skyrmioniums are
mixed together at $j = 0.75 \times 10^{9}\text{ A/m}^{2}$
in Fig.~\ref{Fig9}(b),
in the LS at $j = 2.25 \times 10^{9}\text{ A/m}^{2}$ in Fig.~\ref{Fig9}(c),
and in the MSk state at $j = 3.75 \times 10^{9}\text{ A/m}^{2}$
in Fig.~\ref{Fig9}(d), where all
of the skyrmioniums have been transformed into skyrmions and
the entire system flows along
the intrinsic Hall angle of $\theta_{\rm sk}^{\rm int}=67^{\circ}$.

\begin{figure}
\centering
\includegraphics[width=0.8\columnwidth]{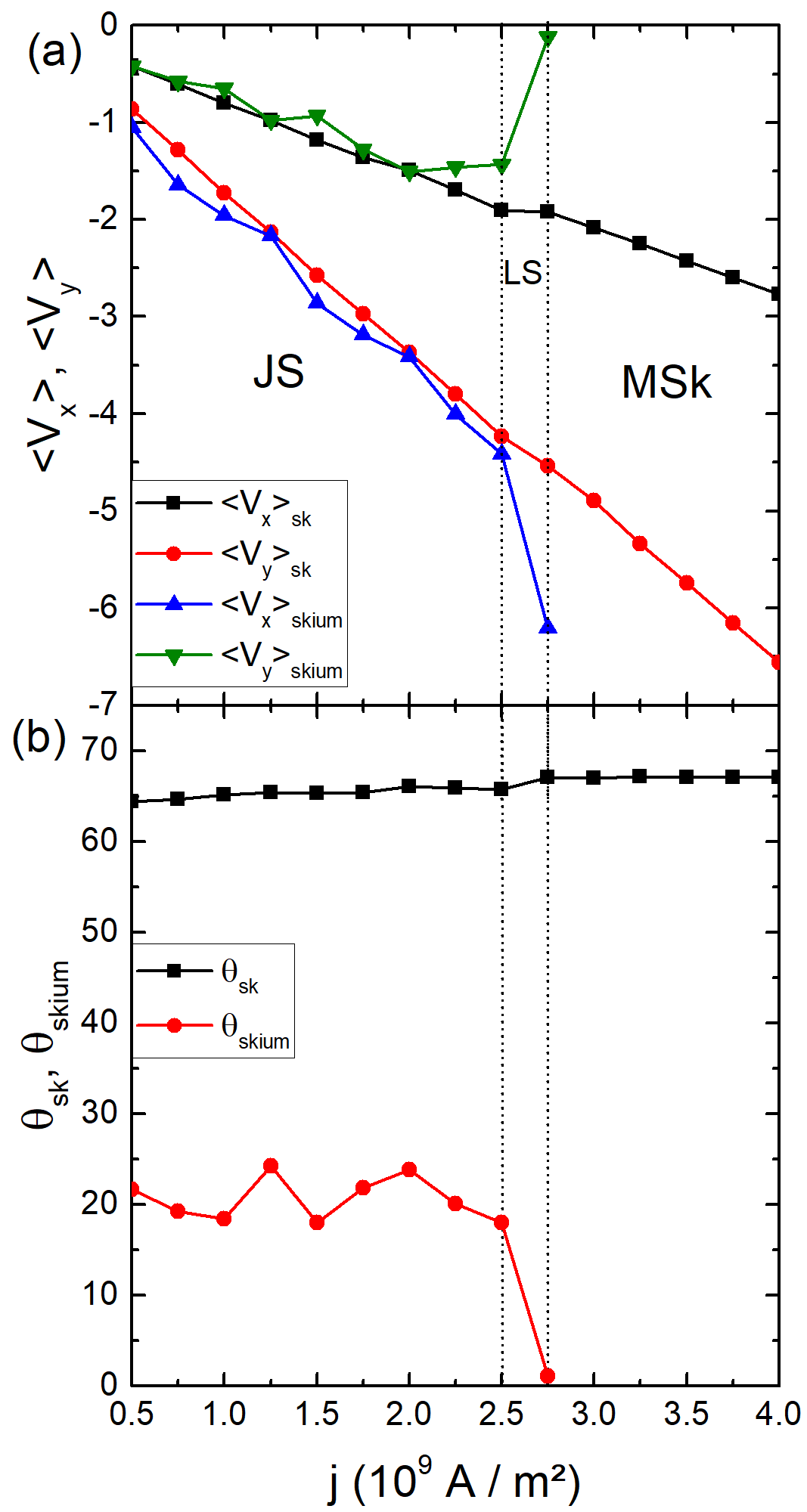}
\caption{
 (a) The velocity components $\langle V_x\rangle_{\rm skium,sk}$ and $\langle V_y\rangle_{\rm skium,sk}$
  and (b) the corresponding angle of motion
  $\theta_{\rm skium,sk}$ with respect to the external drive $j$
  plotted vs $j$
  for the $N_{\rm skium}=2$ and $N_{\rm sk}=26$ sample
  illustrated in Fig.~\ref{Fig11}(a).
  The dashed vertical lines separate the
  partially jammed state (JS), laning state (LS),
 and moving skyrmion lattice (MSk) state.
 }
   \label{Fig10}
\end{figure}

\begin{figure}
\centering
\includegraphics[width=1.0\columnwidth]{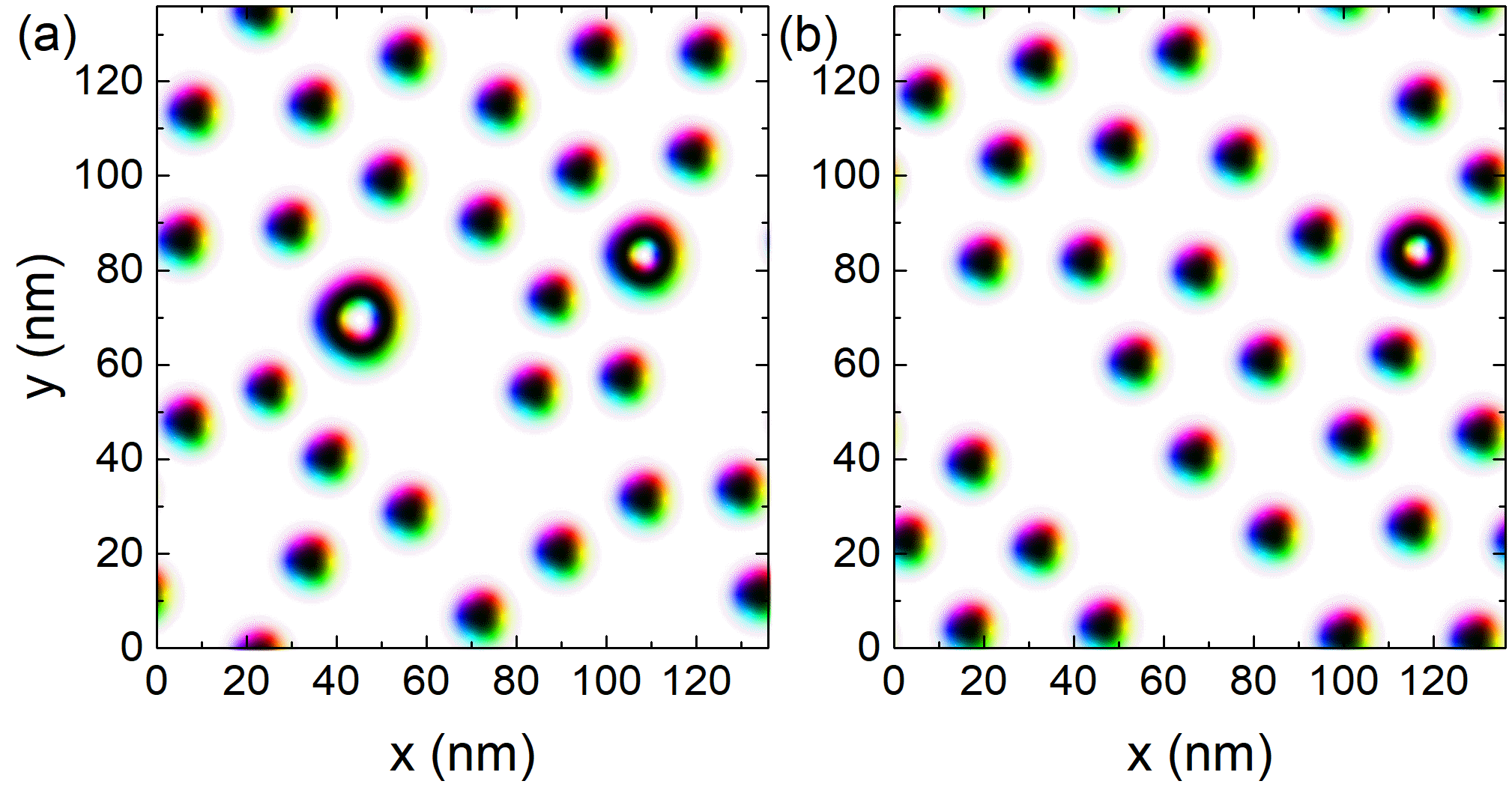}
\caption{
  Snapshots of a system with an initial state of $N_{\rm sk}=26$ and
  $N_{\rm skium}=2$.
  (a) The partially jammed state (JS) at $j = 1.50 \times 10^{9} \text{ A/m}^{2}$,
  where the number of skyrmions and skyrmioniums is unchanged.
  (b) The laning state (LS) at $j = 2.75 \times 10^{9} \text{ A/m}^{2}$,
  where there are now 27 skyrmions and 1 skyrmionium.
The magnetic textures are visualized by projecting the magnetic moments onto the $xy$ plane.
}
\label{Fig11}
\end{figure}

In Fig.~\ref{Fig10} we plot $\left\langle V_x\right\rangle_{\rm sk,skium}$,
$\left\langle V_y\right\rangle_{\rm sk,skium}$, and $\theta_{\rm sk,skium}$
versus $j$ for a system with $N_{\rm skium}=2$ and
$N_{\rm sk}=26$.
The stability region of the JS is increased relative to what we
have shown up to this point, and extends over
$j < 2.5 \times 10^{9} \text{ A/m}^{2}$.
The LS is only stable over a narrow range of
currents, $2.5\times 10^{9} \text{ A/m}^{2} < j \leq 2.75\times 10^{9} \text{ A/m}^{2}$. For $j > 2.75\times 10^{9} \text{ A/m}^{2}$,
all of the skyrmioniums annihilate and the system enters the MSk phase.
Two skyrmioniums are present in the sample
throughout the entire JS, as illustrated in 
Fig.~\ref{Fig11}(a) at $j= 1.50\times 10^{9} \text{ A/m}^{2}$.
That is, both of the skyrmioniums have become jammed
among the skyrmions and are not able to create a
path for phase separation and lane formation. 
When $j > 2.50\times 10^{9} \text{ A/m}^{2}$,
however, one of the skyrmioniums transforms into a skyrmion, reducing the
total effective density of the sample slightly and enabling the
skyrmions to compress enough that the remaining skyrmionium can establish
laning flow, resulting in the formation of
the LS shown in Fig.~\ref{Fig10}.
A snapshot of the LS at $j = 2.75\times 10^{9} \text{ A/m}^{2}$ appears
in Fig.~\ref{Fig11}(b),
where the sample contains 27 skyrmions and only one skyrmionium.
Upon further increasing the drive,
the remaining skyrmionium annihilates and only skyrmions remain in the sample,
very similar to the snapshot shown in Fig.~\ref{Fig9}(d).

Combining
multiple skyrmioniums with multiple skyrmions produces
a much more complex behavior than when there is only a single skyrmionium
present.
When the skyrmion density is not too high,
multiple skyrmioniums can cooperate 
with each other to form lanes and expand the window in which the LS
is stable.
Due to the differences in sizes and dynamics
between the skyrmions and the skyrmioniums,
when the skyrmion density is high,
there is not enough free space available for
multiple skyrmioniums to create a path, and the JS is stabilized
over a wide range of $j$ values.
The formation of lanes becomes possible
only when a portion of the skyrmioniums annihilate and
create more free space for lane formation.

\section{More skyrmioniums than skyrmions}

\begin{figure}
\centering
\includegraphics[width=1.0\columnwidth]{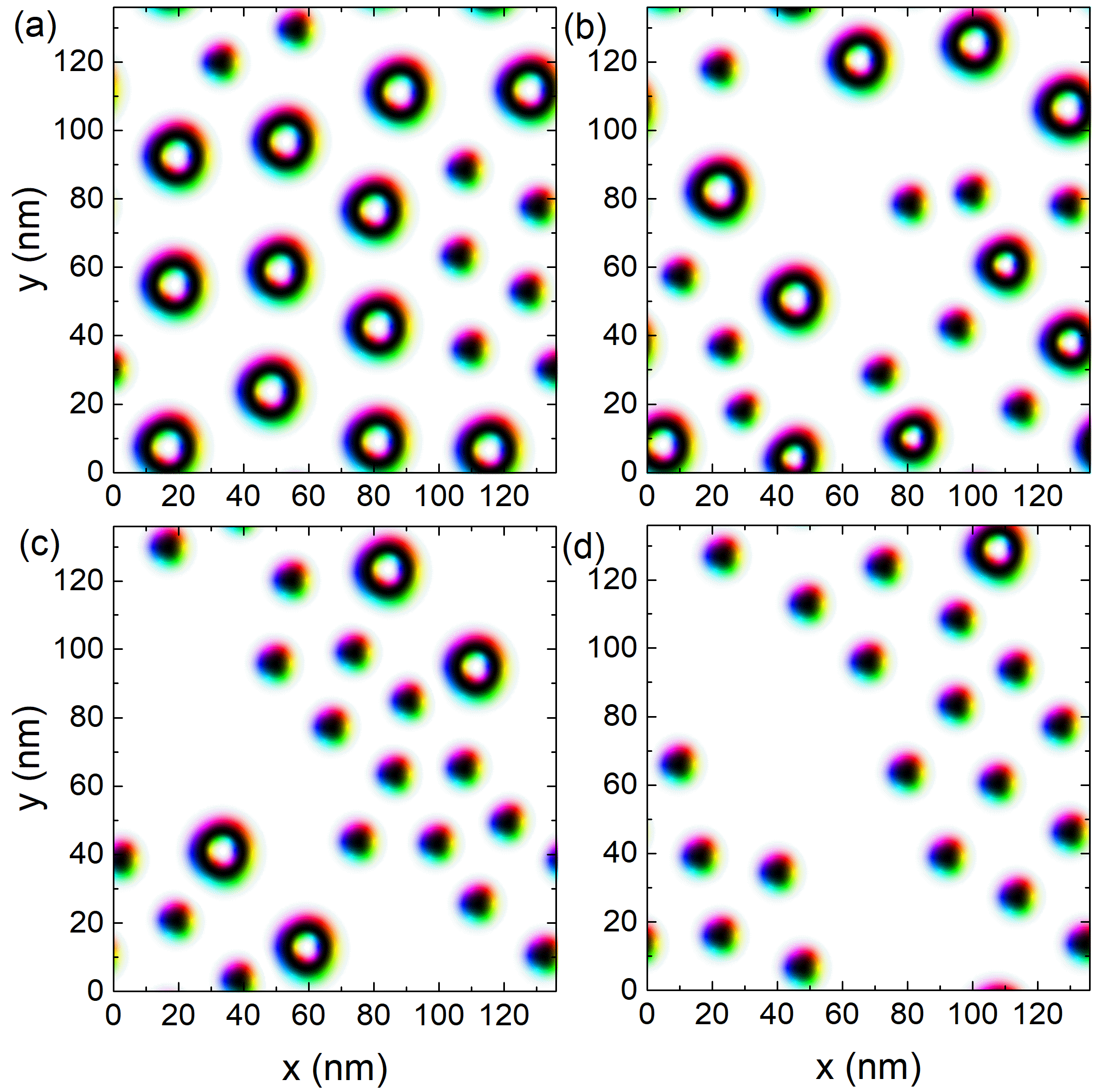}
\caption{
Snapshots of a system with an initial state of
$N_{\rm sk}=8$ and $N_{\rm skium}=12$.
(a) The ground state at $j=0 \text{ A/m}^{2}$.
(b) The jammed state (JS) at $j = 1.00\times 10^{9} \text{ A/m}^{2}$,
where half of the textures are skyrmions 
and the other half are skyrmioniums.
(c) The laning state (LS) at $j = 2.00\times 10^{9} \text{ A/m}^{2}$,
where there are 4 skyrmioniums and 16 skyrmions.
(d) The laning state (LS) at $j = 3.00 \times 10^{9} \text{ A/m}^{2}$,
where there is a single skyrmionium and 19 skyrmions.
The magnetic textures are visualized by projecting the magnetic moments onto the $xy$ plane.
}
\label{Fig12}
\end{figure}

We have shown that when
the number of skyrmions is larger than the number of skyrmioniums,
phase separation of the two species can occur in the form of the
emergence of a laning state, and that this process
depends strongly on the skyrmion density.
In this section we investigate what happens when the ferromagnetic film
contains more skyrmioniums than skyrmions by
preparing a sample with $N_{\rm sk}=8$ and $N_{\rm skium}=12$, as
illustrated in Fig.~\ref{Fig12}(a).

\begin{figure}
 \centering
 \includegraphics[width=0.8\columnwidth]{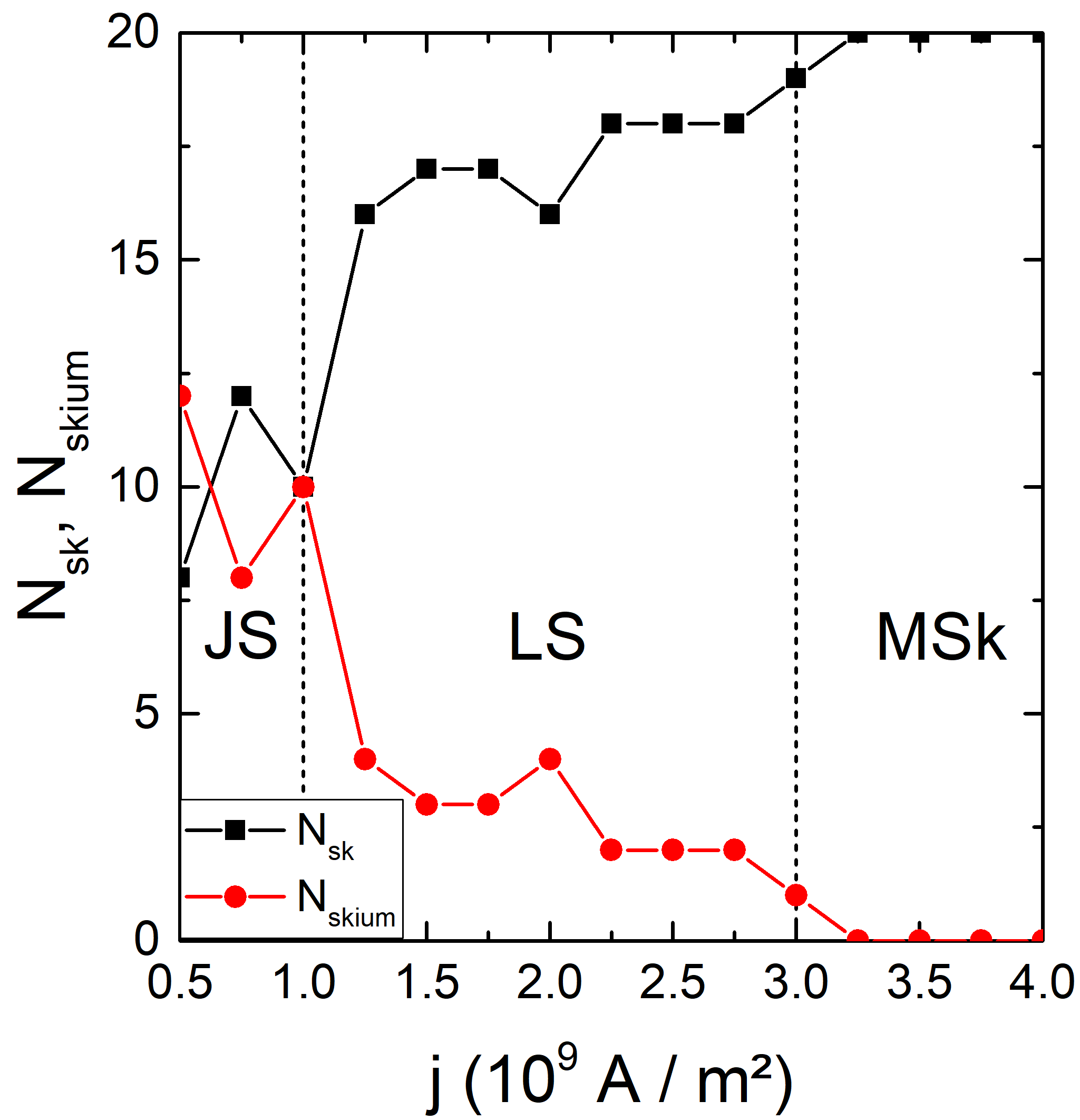}
 \caption{
 The number of skyrmions, $N_{\rm sk}$, and skyrmioniums, $N_{\rm skium}$, present in the sample as a function of the external driving current $j$ for the system from Fig.~\ref{Fig12} that was initialized with $N_{\rm sk}=8$ and $N_{\rm skium}=12.$
The vertical dashed lines separate the jammed state (JS), laning state (LS), and moving skyrmion lattice (MSk).
}
\label{Fig13}
\end{figure}

In Fig. \ref{Fig13}, the plot of $N_{\rm sk}$ and $N_{\rm skium}$ versus
$j$ shows that
skyrmioniums are progressively being collapsed into skyrmions
as the driving current $j$ increases.
At low drives of $j \leq 1.00 \times 10^{9} \text{ A/m}^{2}$,
there is not enough compression of the skyrmioniums
by the current and collisions with other textures to collapse
the skyrmioniums into skyrmions.
In this low drive regime, a JS appears in which
phase separation between skyrmions and skyrmioniums does not occur.
Note that the number of skyrmions and skyrmioniums
present may change depending on the precise value of the current,
but in general about half of the textures are skyrmioniums and
the other half are skyrmions.
For intermediate drives, $1.00 \times 10^{9} \text{ A/m}^{2} < j \leq 3.00 \times 10^{9} \text{ A/m}^{2}$, a LS forms.
At these values of $j$, a portion of
the skyrmioniums collapse into skyrmions, which provides enough
free space
for lane formation and phase separation to occur.
As the drive increases, the number of skyrmioniums continues to fall
and the number of skyrmions continues to increase.
It is important to note that the ground state in Fig.~\ref{Fig12}(a)
is densely packed since the repulsive interactions between skyrmionium pairs
is larger than that of skyrmion pairs
\cite{souza_comparing_2024},
making a skyrmionium array effectively more dense than a skyrmion array
containing the same number of magnetic textures.
Thus, when skyrmioniums are replace by skyrmions at the higher drives,
this effectively creates more space for lane formation.
For high drives of $j > 3.00 \times 10^{9} \text{ A/m}^{2}$,
the current is strong enough to annihilate all of the skyrmioniums
present in the sample, leading to the formation of the MSk phase.

\section{Discussion}

Our atomistic simulations allow us to investigate
the conditions under which skyrmioniums and skyrmions can be stabilized 
in ferromagnetic thin films with a perpendicular applied magnetic field at
zero temperature. We transport
the magnetic textures via spin-orbit torque (SOT);
however, it should be possible to transport skyrmions and skyrmioniums
via spin-transfer-torque (STT) with non-adiabatic terms.
A STT current without a non-adiabatic term
can only transport the skyrmions but would
not transport the skyrmioniums.
In this case, the moving skyrmions would drag the
passive skyrmioniums through the sample,
which would probably cause the 
skyrmioniums to collapse into skyrmions beginning
at relatively low currents.
A STT current with non-adiabatic terms would move both textures at
the same velocity but
along different directions \cite{zhang_control_2016}.
This would be an interesting topic to address in future work,
since lane formation might still occur,
but the stability of the lanes might be affected.
Other alternative methods for transporting the textures
could also be investigated, such as spin waves,
magnetic fields, or temperature gradients 
\cite{lin_particle_2013,raimondo_temperature-gradient-driven_2022,zhang_manipulation_2018,li_dynamics_2018}.
We have not considered thermal effects in our simulations;
however, we expect that mild temperature effects could facilitate the
penetration of the skyrmionium into the skyrmion lattice and enhance the phase separation. Still, since the skyrmioniums are less stable than the
skyrmions, higher temperatures could lead to the
premature annihilation of the inner skyrmion of
the skyrmionium \cite{jiang_transformation_2024}.

\section{Summary}

We have investigated the dynamical phase separation and laning transitions 
of driven skyrmion–skyrmionium mixtures using atomistic simulations. 
Skyrmions move at a finite Hall angle, while skyrmioniums exhibit no Hall
effect and move at twice the speed of the skyrmions.
For a single skyrmionium in a bath of skyrmions,
we observe a low-drive, partially jammed phase in which
the skyrmions
induce 
a finite angle of motion for the skyrmionium
motion due to an effective drag effect.
At higher drives, the system transitions into a laned state,
with the lanes tilted opposite to the Hall angle of the skyrmions.
This results in a reversal of the skyrmionium angle of motion,
as the motion of the
skyrmionium is partially guided by the tilted lane structure.
When the skyrmion density is increased,
we find a high drive state in which the skyrmionium collapses into a skyrmion.
For mixtures containing multiple skyrmioniums,
we observe a partially jammed state in which the skyrmioniums acquire
a finite angle of motion along the direction of the skyrmion
motion,
as well as a laned state where a
finite skyrmionium angle of motion persists.
At sufficiently high drives,
the skyrmioniums once again collapse into skyrmions.
For certain ratios of skyrmioniums to skyrmions,
we find more strongly jammed states in which
the skyrmions and skyrmioniums move together as a rigid cluster
with suppressed laning.
Above a critical skyrmion density, 
the system transitions directly into a fully jammed skyrmion state
that contains no skyrmioniums.
Unlike what is observed in colloidal systems,
the lanes in our system are tilted relative to the driving direction,
similar to the
tilted lanes recently observed
for pedestrian crowds moving in opposite directions
with chiral rules of motion.
In the skyrmion/skyrmionium system,
the chirality arises from the finite skyrmion Hall angle.
We map out a
dynamic phase diagram as a function of current and skyrmion density.
Although our results focus on skyrmion–skyrmionium mixtures,
we expect similar laning transitions to occur
for mixtures of other magnetic textures
that have different Hall angles or mobilities.
Our findings indicate that mixtures of magnetic textures with distinct
topologies represent
a promising and fruitful
class of systems for exploring laning transitions and collective dynamics.

\acknowledgments
This work was supported by the US Department of Energy through
the Los Alamos National Laboratory.  
Los Alamos National Laboratory is
operated by Triad National Security, LLC, for the National Nuclear Security
Administration of the U. S. Department of Energy (Contract No. 892333218NCA000001).
N.P.V. acknowledges
funding from
Funda\c{c}\~{a}o de Amparo \`{a} Pesquisa do Estado de S\~{a}o Paulo - FAPESP (Grant 2024/13248-5).
J.C.B.S. acknowledges
funding from
Funda\c{c}\~{a}o de Amparo \`{a} Pesquisa do Estado de S\~{a}o Paulo - FAPESP (Grant 2021/04941-0).
We would like to thank FAPESP for providing the computational resources used in this
work (Grant: 2024/02941-1).

\bibliography{mybib}

\begin{thebibliography}{66}%
\makeatletter
\providecommand \@ifxundefined [1]{%
 \@ifx{#1\undefined}
}%
\providecommand \@ifnum [1]{%
 \ifnum #1\expandafter \@firstoftwo
 \else \expandafter \@secondoftwo
 \fi
}%
\providecommand \@ifx [1]{%
 \ifx #1\expandafter \@firstoftwo
 \else \expandafter \@secondoftwo
 \fi
}%
\providecommand \natexlab [1]{#1}%
\providecommand \enquote  [1]{``#1''}%
\providecommand \bibnamefont  [1]{#1}%
\providecommand \bibfnamefont [1]{#1}%
\providecommand \citenamefont [1]{#1}%
\providecommand \href@noop [0]{\@secondoftwo}%
\providecommand \href [0]{\begingroup \@sanitize@url \@href}%
\providecommand \@href[1]{\@@startlink{#1}\@@href}%
\providecommand \@@href[1]{\endgroup#1\@@endlink}%
\providecommand \@sanitize@url [0]{\catcode `\\12\catcode `\$12\catcode `\&12\catcode `\#12\catcode `\^12\catcode `\_12\catcode `\%12\relax}%
\providecommand \@@startlink[1]{}%
\providecommand \@@endlink[0]{}%
\providecommand \url  [0]{\begingroup\@sanitize@url \@url }%
\providecommand \@url [1]{\endgroup\@href {#1}{\urlprefix }}%
\providecommand \urlprefix  [0]{URL }%
\providecommand \Eprint [0]{\href }%
\providecommand \doibase [0]{http://dx.doi.org/}%
\providecommand \selectlanguage [0]{\@gobble}%
\providecommand \bibinfo  [0]{\@secondoftwo}%
\providecommand \bibfield  [0]{\@secondoftwo}%
\providecommand \translation [1]{[#1]}%
\providecommand \BibitemOpen [0]{}%
\providecommand \bibitemStop [0]{}%
\providecommand \bibitemNoStop [0]{.\EOS\space}%
\providecommand \EOS [0]{\spacefactor3000\relax}%
\providecommand \BibitemShut  [1]{\csname bibitem#1\endcsname}%
\let\auto@bib@innerbib\@empty
\bibitem [{\citenamefont {Sadr-Lahijany}\ \emph {et~al.}(1997)\citenamefont {Sadr-Lahijany}, \citenamefont {Ray},\ and\ \citenamefont {Stanley}}]{sadr-lahijany_dispersity-driven_1997}%
  \BibitemOpen
  \bibfield  {author} {\bibinfo {author} {\bibfnamefont {M.~R.}\ \bibnamefont {Sadr-Lahijany}}, \bibinfo {author} {\bibfnamefont {P.}~\bibnamefont {Ray}}, \ and\ \bibinfo {author} {\bibfnamefont {H.~E.}\ \bibnamefont {Stanley}},\ }\bibfield  {title} {\enquote {\bibinfo {title} {Dispersity-driven melting transition in two-dimensional solids},}\ }\href {\doibase 10.1103/PhysRevLett.79.3206} {\bibfield  {journal} {\bibinfo  {journal} {Phys. Rev. Lett.}\ }\textbf {\bibinfo {volume} {79}},\ \bibinfo {pages} {3206--3209} (\bibinfo {year} {1997})}\BibitemShut {NoStop}%
\bibitem [{\citenamefont {Hamanaka}\ and\ \citenamefont {Onuki}(2006)}]{hamanaka_transitions_2006}%
  \BibitemOpen
  \bibfield  {author} {\bibinfo {author} {\bibfnamefont {T.}~\bibnamefont {Hamanaka}}\ and\ \bibinfo {author} {\bibfnamefont {A.}~\bibnamefont {Onuki}},\ }\bibfield  {title} {\enquote {\bibinfo {title} {Transitions among crystal, glass, and liquid in a binary mixture with changing particle-size ratio and temperature},}\ }\href {\doibase 10.1103/PhysRevE.74.011506} {\bibfield  {journal} {\bibinfo  {journal} {Phys. Rev. E}\ }\textbf {\bibinfo {volume} {74}},\ \bibinfo {pages} {011506} (\bibinfo {year} {2006})}\BibitemShut {NoStop}%
\bibitem [{\citenamefont {Reichhardt}\ and\ \citenamefont {Reichhardt}(2008)}]{reichhardt_disordering_2008}%
  \BibitemOpen
  \bibfield  {author} {\bibinfo {author} {\bibfnamefont {C.}~\bibnamefont {Reichhardt}}\ and\ \bibinfo {author} {\bibfnamefont {C.~J.~Olson}\ \bibnamefont {Reichhardt}},\ }\bibfield  {title} {\enquote {\bibinfo {title} {Disordering transitions and peak effect in polydisperse particle systems},}\ }\href {\doibase 10.1103/PhysRevE.77.041401} {\bibfield  {journal} {\bibinfo  {journal} {Phys. Rev. E}\ }\textbf {\bibinfo {volume} {77}},\ \bibinfo {pages} {041401} (\bibinfo {year} {2008})}\BibitemShut {NoStop}%
\bibitem [{\citenamefont {Schmittmann}\ and\ \citenamefont {Zia}(1998)}]{schmittmann_driven_1998}%
  \BibitemOpen
  \bibfield  {author} {\bibinfo {author} {\bibfnamefont {B.}~\bibnamefont {Schmittmann}}\ and\ \bibinfo {author} {\bibfnamefont {R.~K.~P.}\ \bibnamefont {Zia}},\ }\bibfield  {title} {\enquote {\bibinfo {title} {Driven diffusive systems. {A}n introduction and recent developments},}\ }\href {\doibase 10.1016/S0370-1573(98)00005-2} {\bibfield  {journal} {\bibinfo  {journal} {Phys. Rep.}\ }\textbf {\bibinfo {volume} {301}},\ \bibinfo {pages} {45--64} (\bibinfo {year} {1998})}\BibitemShut {NoStop}%
\bibitem [{\citenamefont {Reichhardt}\ and\ \citenamefont {Reichhardt}(2019)}]{reichhardt_disordering_2019}%
  \BibitemOpen
  \bibfield  {author} {\bibinfo {author} {\bibfnamefont {C.~J.~O.}\ \bibnamefont {Reichhardt}}\ and\ \bibinfo {author} {\bibfnamefont {C.}~\bibnamefont {Reichhardt}},\ }\bibfield  {title} {\enquote {\bibinfo {title} {Disordering, clustering, and laning transitions in particle systems with dispersion in the {M}agnus term},}\ }\href {\doibase 10.1103/PhysRevE.99.012606} {\bibfield  {journal} {\bibinfo  {journal} {Phys. Rev. E}\ }\textbf {\bibinfo {volume} {99}},\ \bibinfo {pages} {012606} (\bibinfo {year} {2019})}\BibitemShut {NoStop}%
\bibitem [{\citenamefont {Helbing}\ \emph {et~al.}(2000)\citenamefont {Helbing}, \citenamefont {Farkas},\ and\ \citenamefont {Vicsek}}]{helbing_freezing_2000}%
  \BibitemOpen
  \bibfield  {author} {\bibinfo {author} {\bibfnamefont {D.}~\bibnamefont {Helbing}}, \bibinfo {author} {\bibfnamefont {I.~J.}\ \bibnamefont {Farkas}}, \ and\ \bibinfo {author} {\bibfnamefont {T.}~\bibnamefont {Vicsek}},\ }\bibfield  {title} {\enquote {\bibinfo {title} {Freezing by heating in a driven mesoscopic system},}\ }\href {\doibase 10.1103/PhysRevLett.84.1240} {\bibfield  {journal} {\bibinfo  {journal} {Phys. Rev. Lett.}\ }\textbf {\bibinfo {volume} {84}},\ \bibinfo {pages} {1240--1243} (\bibinfo {year} {2000})}\BibitemShut {NoStop}%
\bibitem [{\citenamefont {Dzubiella}\ \emph {et~al.}(2002)\citenamefont {Dzubiella}, \citenamefont {Hoffmann},\ and\ \citenamefont {L\"owen}}]{dzubiella_lane_2002}%
  \BibitemOpen
  \bibfield  {author} {\bibinfo {author} {\bibfnamefont {J.}~\bibnamefont {Dzubiella}}, \bibinfo {author} {\bibfnamefont {G.~P.}\ \bibnamefont {Hoffmann}}, \ and\ \bibinfo {author} {\bibfnamefont {H.}~\bibnamefont {L\"owen}},\ }\bibfield  {title} {\enquote {\bibinfo {title} {Lane formation in colloidal mixtures driven by an external field},}\ }\href {\doibase 10.1103/PhysRevE.65.021402} {\bibfield  {journal} {\bibinfo  {journal} {Phys. Rev. E}\ }\textbf {\bibinfo {volume} {65}},\ \bibinfo {pages} {021402} (\bibinfo {year} {2002})}\BibitemShut {NoStop}%
\bibitem [{\citenamefont {Netz}(2003)}]{netz_conduction_2003}%
  \BibitemOpen
  \bibfield  {author} {\bibinfo {author} {\bibfnamefont {R.~R.}\ \bibnamefont {Netz}},\ }\bibfield  {title} {\enquote {\bibinfo {title} {Conduction and diffusion in two-dimensional electrolytes},}\ }\href {\doibase 10.1209/epl/i2003-00557-x} {\bibfield  {journal} {\bibinfo  {journal} {Europhys. Lett.}\ }\textbf {\bibinfo {volume} {63}},\ \bibinfo {pages} {616--622} (\bibinfo {year} {2003})}\BibitemShut {NoStop}%
\bibitem [{\citenamefont {Ikeda}\ \emph {et~al.}(2012)\citenamefont {Ikeda}, \citenamefont {Wada},\ and\ \citenamefont {Hayakawa}}]{ikeda_instabilities_2012}%
  \BibitemOpen
  \bibfield  {author} {\bibinfo {author} {\bibfnamefont {M.}~\bibnamefont {Ikeda}}, \bibinfo {author} {\bibfnamefont {H.}~\bibnamefont {Wada}}, \ and\ \bibinfo {author} {\bibfnamefont {H.}~\bibnamefont {Hayakawa}},\ }\bibfield  {title} {\enquote {\bibinfo {title} {Instabilities and turbulence-like dynamics in an oppositely driven binary particle mixture},}\ }\href {\doibase 10.1209/0295-5075/99/68005} {\bibfield  {journal} {\bibinfo  {journal} {EPL}\ }\textbf {\bibinfo {volume} {99}},\ \bibinfo {pages} {68005} (\bibinfo {year} {2012})}\BibitemShut {NoStop}%
\bibitem [{\citenamefont {W\"achtler}\ \emph {et~al.}(2016)\citenamefont {W\"achtler}, \citenamefont {Kogler},\ and\ \citenamefont {Klapp}}]{wachtler_lane_2016}%
  \BibitemOpen
  \bibfield  {author} {\bibinfo {author} {\bibfnamefont {C.~W.}\ \bibnamefont {W\"achtler}}, \bibinfo {author} {\bibfnamefont {F.}~\bibnamefont {Kogler}}, \ and\ \bibinfo {author} {\bibfnamefont {S.~H.~L.}\ \bibnamefont {Klapp}},\ }\bibfield  {title} {\enquote {\bibinfo {title} {Lane formation in a driven attractive fluid},}\ }\href {\doibase 10.1103/PhysRevE.94.052603} {\bibfield  {journal} {\bibinfo  {journal} {Phys. Rev. E}\ }\textbf {\bibinfo {volume} {94}},\ \bibinfo {pages} {052603} (\bibinfo {year} {2016})}\BibitemShut {NoStop}%
\bibitem [{\citenamefont {Poncet}\ \emph {et~al.}(2017)\citenamefont {Poncet}, \citenamefont {B\'enichou}, \citenamefont {D\'emery},\ and\ \citenamefont {Oshanin}}]{poncet_universal_2017}%
  \BibitemOpen
  \bibfield  {author} {\bibinfo {author} {\bibfnamefont {A.}~\bibnamefont {Poncet}}, \bibinfo {author} {\bibfnamefont {O.}~\bibnamefont {B\'enichou}}, \bibinfo {author} {\bibfnamefont {V.}~\bibnamefont {D\'emery}}, \ and\ \bibinfo {author} {\bibfnamefont {G.}~\bibnamefont {Oshanin}},\ }\bibfield  {title} {\enquote {\bibinfo {title} {Universal long ranged correlations in driven binary mixtures},}\ }\href {\doibase 10.1103/PhysRevLett.118.118002} {\bibfield  {journal} {\bibinfo  {journal} {Phys. Rev. Lett.}\ }\textbf {\bibinfo {volume} {118}},\ \bibinfo {pages} {118002} (\bibinfo {year} {2017})}\BibitemShut {NoStop}%
\bibitem [{\citenamefont {Bain}\ and\ \citenamefont {Bartolo}(2017)}]{bain_critical_2017}%
  \BibitemOpen
  \bibfield  {author} {\bibinfo {author} {\bibfnamefont {N.}~\bibnamefont {Bain}}\ and\ \bibinfo {author} {\bibfnamefont {D.}~\bibnamefont {Bartolo}},\ }\bibfield  {title} {\enquote {\bibinfo {title} {Critical mingling and universal correlations in model binary active liquids},}\ }\href {\doibase 10.1038/ncomms15969} {\bibfield  {journal} {\bibinfo  {journal} {Nature Commun.}\ }\textbf {\bibinfo {volume} {8}},\ \bibinfo {pages} {15969} (\bibinfo {year} {2017})}\BibitemShut {NoStop}%
\bibitem [{\citenamefont {Reichhardt}\ \emph {et~al.}(2018)\citenamefont {Reichhardt}, \citenamefont {Thibault}, \citenamefont {Papanikolaou},\ and\ \citenamefont {Reichhardt}}]{reichhardt_laning_2018}%
  \BibitemOpen
  \bibfield  {author} {\bibinfo {author} {\bibfnamefont {C.}~\bibnamefont {Reichhardt}}, \bibinfo {author} {\bibfnamefont {J.}~\bibnamefont {Thibault}}, \bibinfo {author} {\bibfnamefont {S.}~\bibnamefont {Papanikolaou}}, \ and\ \bibinfo {author} {\bibfnamefont {C.~J.~O.}\ \bibnamefont {Reichhardt}},\ }\bibfield  {title} {\enquote {\bibinfo {title} {Laning and clustering transitions in driven binary active matter systems},}\ }\href {\doibase 10.1103/PhysRevE.98.022603} {\bibfield  {journal} {\bibinfo  {journal} {Phys. Rev. E}\ }\textbf {\bibinfo {volume} {98}},\ \bibinfo {pages} {022603} (\bibinfo {year} {2018})}\BibitemShut {NoStop}%
\bibitem [{\citenamefont {Bacik}\ \emph {et~al.}(2023)\citenamefont {Bacik}, \citenamefont {Bacik},\ and\ \citenamefont {Rogers}}]{bacik_lane_2023}%
  \BibitemOpen
  \bibfield  {author} {\bibinfo {author} {\bibfnamefont {K.~A.}\ \bibnamefont {Bacik}}, \bibinfo {author} {\bibfnamefont {B.~S.}\ \bibnamefont {Bacik}}, \ and\ \bibinfo {author} {\bibfnamefont {T.}~\bibnamefont {Rogers}},\ }\bibfield  {title} {\enquote {\bibinfo {title} {Lane nucleation in complex active flows},}\ }\href {\doibase 10.1126/science.add8091} {\bibfield  {journal} {\bibinfo  {journal} {Science}\ }\textbf {\bibinfo {volume} {379}},\ \bibinfo {pages} {923--928} (\bibinfo {year} {2023})}\BibitemShut {NoStop}%
\bibitem [{\citenamefont {M{\" u}hlbauer}\ \emph {et~al.}(2009)\citenamefont {M{\" u}hlbauer}, \citenamefont {Binz}, \citenamefont {Jonietz}, \citenamefont {Pfleiderer}, \citenamefont {Rosch}, \citenamefont {Neubauer}, \citenamefont {Georgii},\ and\ \citenamefont {B{\" o}ni}}]{muhlbauer_skyrmion_2009}%
  \BibitemOpen
  \bibfield  {author} {\bibinfo {author} {\bibfnamefont {S.}~\bibnamefont {M{\" u}hlbauer}}, \bibinfo {author} {\bibfnamefont {B.}~\bibnamefont {Binz}}, \bibinfo {author} {\bibfnamefont {F.}~\bibnamefont {Jonietz}}, \bibinfo {author} {\bibfnamefont {C.}~\bibnamefont {Pfleiderer}}, \bibinfo {author} {\bibfnamefont {A.}~\bibnamefont {Rosch}}, \bibinfo {author} {\bibfnamefont {A.}~\bibnamefont {Neubauer}}, \bibinfo {author} {\bibfnamefont {R.}~\bibnamefont {Georgii}}, \ and\ \bibinfo {author} {\bibfnamefont {P.}~\bibnamefont {B{\" o}ni}},\ }\bibfield  {title} {\enquote {\bibinfo {title} {Skyrmion lattice in a chiral magnet},}\ }\href {\doibase 10.1126/science.1166767} {\bibfield  {journal} {\bibinfo  {journal} {Science}\ }\textbf {\bibinfo {volume} {323}},\ \bibinfo {pages} {915--919} (\bibinfo {year} {2009})}\BibitemShut {NoStop}%
\bibitem [{\citenamefont {Fert}\ \emph {et~al.}(2017)\citenamefont {Fert}, \citenamefont {Reyren},\ and\ \citenamefont {Cros}}]{fert_magnetic_2017}%
  \BibitemOpen
  \bibfield  {author} {\bibinfo {author} {\bibfnamefont {A.}~\bibnamefont {Fert}}, \bibinfo {author} {\bibfnamefont {N.}~\bibnamefont {Reyren}}, \ and\ \bibinfo {author} {\bibfnamefont {V.}~\bibnamefont {Cros}},\ }\bibfield  {title} {\enquote {\bibinfo {title} {Magnetic skyrmions: advances in physics and potential applications},}\ }\href {\doibase 10.1038/natrevmats.2017.31} {\bibfield  {journal} {\bibinfo  {journal} {Nature Rev. Mater.}\ }\textbf {\bibinfo {volume} {2}},\ \bibinfo {pages} {17031} (\bibinfo {year} {2017})}\BibitemShut {NoStop}%
\bibitem [{\citenamefont {Bogdanov}\ and\ \citenamefont {Hubert}(1994)}]{bogdanov_thermodynamically_1994}%
  \BibitemOpen
  \bibfield  {author} {\bibinfo {author} {\bibfnamefont {A.}~\bibnamefont {Bogdanov}}\ and\ \bibinfo {author} {\bibfnamefont {A.}~\bibnamefont {Hubert}},\ }\bibfield  {title} {\enquote {\bibinfo {title} {Thermodynamically stable magnetic vortex states in magnetic crystals},}\ }\href {\doibase 10.1016/0304-8853(94)90046-9} {\bibfield  {journal} {\bibinfo  {journal} {J. Mag. Mag. Mater.}\ }\textbf {\bibinfo {volume} {138}},\ \bibinfo {pages} {255--269} (\bibinfo {year} {1994})}\BibitemShut {NoStop}%
\bibitem [{\citenamefont {R{\" o}{\ss}ler}\ \emph {et~al.}(2006)\citenamefont {R{\" o}{\ss}ler}, \citenamefont {Bogdanov},\ and\ \citenamefont {Pfleiderer}}]{rosler_spontaneous_2006}%
  \BibitemOpen
  \bibfield  {author} {\bibinfo {author} {\bibfnamefont {U.~K.}\ \bibnamefont {R{\" o}{\ss}ler}}, \bibinfo {author} {\bibfnamefont {A.~N.}\ \bibnamefont {Bogdanov}}, \ and\ \bibinfo {author} {\bibfnamefont {C.}~\bibnamefont {Pfleiderer}},\ }\bibfield  {title} {\enquote {\bibinfo {title} {Spontaneous skyrmion ground states in magnetic metals},}\ }\href {\doibase 10.1038/nature05056} {\bibfield  {journal} {\bibinfo  {journal} {Nature (London)}\ }\textbf {\bibinfo {volume} {442}},\ \bibinfo {pages} {797--801} (\bibinfo {year} {2006})}\BibitemShut {NoStop}%
\bibitem [{\citenamefont {Yu}\ \emph {et~al.}(2010)\citenamefont {Yu}, \citenamefont {Onose}, \citenamefont {Kanazawa}, \citenamefont {Park}, \citenamefont {Han}, \citenamefont {Matsui}, \citenamefont {Nagaosa},\ and\ \citenamefont {Tokura}}]{yu_real-space_2010}%
  \BibitemOpen
  \bibfield  {author} {\bibinfo {author} {\bibfnamefont {X.~Z.}\ \bibnamefont {Yu}}, \bibinfo {author} {\bibfnamefont {Y.}~\bibnamefont {Onose}}, \bibinfo {author} {\bibfnamefont {N.}~\bibnamefont {Kanazawa}}, \bibinfo {author} {\bibfnamefont {J.~H.}\ \bibnamefont {Park}}, \bibinfo {author} {\bibfnamefont {J.~H.}\ \bibnamefont {Han}}, \bibinfo {author} {\bibfnamefont {Y.}~\bibnamefont {Matsui}}, \bibinfo {author} {\bibfnamefont {N.}~\bibnamefont {Nagaosa}}, \ and\ \bibinfo {author} {\bibfnamefont {Y.}~\bibnamefont {Tokura}},\ }\bibfield  {title} {\enquote {\bibinfo {title} {Real-space observation of a two-dimensional skyrmion crystal},}\ }\href {\doibase 10.1038/nature09124} {\bibfield  {journal} {\bibinfo  {journal} {Nature (London)}\ }\textbf {\bibinfo {volume} {465}},\ \bibinfo {pages} {901--904} (\bibinfo {year} {2010})}\BibitemShut {NoStop}%
\bibitem [{\citenamefont {Schulz}\ \emph {et~al.}(2012)\citenamefont {Schulz}, \citenamefont {Ritz}, \citenamefont {Bauer}, \citenamefont {Halder}, \citenamefont {Wagner}, \citenamefont {Franz}, \citenamefont {Pfleiderer}, \citenamefont {Everschor}, \citenamefont {Garst},\ and\ \citenamefont {Rosch}}]{schulz_emergent_2012}%
  \BibitemOpen
  \bibfield  {author} {\bibinfo {author} {\bibfnamefont {T.}~\bibnamefont {Schulz}}, \bibinfo {author} {\bibfnamefont {R.}~\bibnamefont {Ritz}}, \bibinfo {author} {\bibfnamefont {A.}~\bibnamefont {Bauer}}, \bibinfo {author} {\bibfnamefont {M.}~\bibnamefont {Halder}}, \bibinfo {author} {\bibfnamefont {M.}~\bibnamefont {Wagner}}, \bibinfo {author} {\bibfnamefont {C.}~\bibnamefont {Franz}}, \bibinfo {author} {\bibfnamefont {C.}~\bibnamefont {Pfleiderer}}, \bibinfo {author} {\bibfnamefont {K.}~\bibnamefont {Everschor}}, \bibinfo {author} {\bibfnamefont {M.}~\bibnamefont {Garst}}, \ and\ \bibinfo {author} {\bibfnamefont {A.}~\bibnamefont {Rosch}},\ }\bibfield  {title} {\enquote {\bibinfo {title} {Emergent electrodynamics of skyrmions in a chiral magnet},}\ }\href {\doibase 10.1038/NPHYS2231} {\bibfield  {journal} {\bibinfo  {journal} {Nature Phys.}\ }\textbf {\bibinfo {volume} {8}},\ \bibinfo {pages} {301--304} (\bibinfo {year} {2012})}\BibitemShut {NoStop}%
\bibitem [{\citenamefont {Jonietz}\ \emph {et~al.}(2010)\citenamefont {Jonietz}, \citenamefont {M{\" u}hlbauer}, \citenamefont {Pfleiderer}, \citenamefont {Neubauer}, \citenamefont {M{\" u}nzer}, \citenamefont {Bauer}, \citenamefont {Adams}, \citenamefont {Georgii}, \citenamefont {B{\" o}ni}, \citenamefont {Duine}, \citenamefont {Everschor}, \citenamefont {Garst},\ and\ \citenamefont {Rosch}}]{jonietz_spin_2010}%
  \BibitemOpen
  \bibfield  {author} {\bibinfo {author} {\bibfnamefont {F.}~\bibnamefont {Jonietz}}, \bibinfo {author} {\bibfnamefont {S.}~\bibnamefont {M{\" u}hlbauer}}, \bibinfo {author} {\bibfnamefont {C.}~\bibnamefont {Pfleiderer}}, \bibinfo {author} {\bibfnamefont {A.}~\bibnamefont {Neubauer}}, \bibinfo {author} {\bibfnamefont {W.}~\bibnamefont {M{\" u}nzer}}, \bibinfo {author} {\bibfnamefont {A.}~\bibnamefont {Bauer}}, \bibinfo {author} {\bibfnamefont {T.}~\bibnamefont {Adams}}, \bibinfo {author} {\bibfnamefont {R.}~\bibnamefont {Georgii}}, \bibinfo {author} {\bibfnamefont {P.}~\bibnamefont {B{\" o}ni}}, \bibinfo {author} {\bibfnamefont {R.~A.}\ \bibnamefont {Duine}}, \bibinfo {author} {\bibfnamefont {K.}~\bibnamefont {Everschor}}, \bibinfo {author} {\bibfnamefont {M.}~\bibnamefont {Garst}}, \ and\ \bibinfo {author} {\bibfnamefont {A.}~\bibnamefont {Rosch}},\ }\bibfield  {title} {\enquote {\bibinfo {title} {Spin transfer torques in {MnSi} at ultralow current densities},}\ }\href {\doibase 10.1126/science.1195709}
  {\bibfield  {journal} {\bibinfo  {journal} {Science}\ }\textbf {\bibinfo {volume} {330}},\ \bibinfo {pages} {1648--1651} (\bibinfo {year} {2010})}\BibitemShut {NoStop}%
\bibitem [{\citenamefont {Reichhardt}\ and\ \citenamefont {Reichhardt}(2017)}]{reichhardt_depinning_2016}%
  \BibitemOpen
  \bibfield  {author} {\bibinfo {author} {\bibfnamefont {C.}~\bibnamefont {Reichhardt}}\ and\ \bibinfo {author} {\bibfnamefont {C.~J.~Olson}\ \bibnamefont {Reichhardt}},\ }\bibfield  {title} {\enquote {\bibinfo {title} {Depinning and nonequilibrium dynamic phases of particle assemblies driven over random and ordered substrates: a review},}\ }\href {\doibase 10.1088/1361-6633/80/2/026501} {\bibfield  {journal} {\bibinfo  {journal} {Rep. Prog. Phys.}\ }\textbf {\bibinfo {volume} {80}},\ \bibinfo {pages} {026501} (\bibinfo {year} {2017})}\BibitemShut {NoStop}%
\bibitem [{\citenamefont {Reichhardt}\ \emph {et~al.}(2022)\citenamefont {Reichhardt}, \citenamefont {Reichhardt},\ and\ \citenamefont {Milo{\v s}evi{\' c}}}]{reichhardt_statics_2022}%
  \BibitemOpen
  \bibfield  {author} {\bibinfo {author} {\bibfnamefont {C.}~\bibnamefont {Reichhardt}}, \bibinfo {author} {\bibfnamefont {C.~J.~O.}\ \bibnamefont {Reichhardt}}, \ and\ \bibinfo {author} {\bibfnamefont {M.}~\bibnamefont {Milo{\v s}evi{\' c}}},\ }\bibfield  {title} {\enquote {\bibinfo {title} {Statics and dynamics of skyrmions interacting with disorder and nanostructures},}\ }\href {\doibase 10.1103/RevModPhys.94.035005} {\bibfield  {journal} {\bibinfo  {journal} {Rev. Mod. Phys.}\ }\textbf {\bibinfo {volume} {94}},\ \bibinfo {pages} {035005} (\bibinfo {year} {2022})}\BibitemShut {NoStop}%
\bibitem [{\citenamefont {Skyrme}(1962)}]{skyrme_unified_1962}%
  \BibitemOpen
  \bibfield  {author} {\bibinfo {author} {\bibfnamefont {T.~H.~R.}\ \bibnamefont {Skyrme}},\ }\bibfield  {title} {\enquote {\bibinfo {title} {A unified field theory of mesons and baryons},}\ }\href {\doibase 10.1016/0029-5582(62)90775-7} {\bibfield  {journal} {\bibinfo  {journal} {Nucl. Phys.}\ }\textbf {\bibinfo {volume} {31}},\ \bibinfo {pages} {556} (\bibinfo {year} {1962})}\BibitemShut {NoStop}%
\bibitem [{\citenamefont {Jiang}\ \emph {et~al.}(2015)\citenamefont {Jiang}, \citenamefont {Upadhyaya}, \citenamefont {Zhang}, \citenamefont {Yu}, \citenamefont {Jungfleisch}, \citenamefont {Fradin}, \citenamefont {Pearson}, \citenamefont {Tserkovnyak}, \citenamefont {Wang}, \citenamefont {Heinonen}, \citenamefont {te~Velthuis},\ and\ \citenamefont {Hoffmann}}]{jiang_blowing_2015}%
  \BibitemOpen
  \bibfield  {author} {\bibinfo {author} {\bibfnamefont {W.}~\bibnamefont {Jiang}}, \bibinfo {author} {\bibfnamefont {P.}~\bibnamefont {Upadhyaya}}, \bibinfo {author} {\bibfnamefont {W.}~\bibnamefont {Zhang}}, \bibinfo {author} {\bibfnamefont {G.}~\bibnamefont {Yu}}, \bibinfo {author} {\bibfnamefont {M.~B.}\ \bibnamefont {Jungfleisch}}, \bibinfo {author} {\bibfnamefont {F.~Y.}\ \bibnamefont {Fradin}}, \bibinfo {author} {\bibfnamefont {J.~E.}\ \bibnamefont {Pearson}}, \bibinfo {author} {\bibfnamefont {Y.}~\bibnamefont {Tserkovnyak}}, \bibinfo {author} {\bibfnamefont {K.~L.}\ \bibnamefont {Wang}}, \bibinfo {author} {\bibfnamefont {O.}~\bibnamefont {Heinonen}}, \bibinfo {author} {\bibfnamefont {S.~G.~E.}\ \bibnamefont {te~Velthuis}}, \ and\ \bibinfo {author} {\bibfnamefont {A.}~\bibnamefont {Hoffmann}},\ }\bibfield  {title} {\enquote {\bibinfo {title} {Blowing magnetic skyrmion bubbles},}\ }\href {\doibase 10.1126/science.aaa1442} {\bibfield  {journal} {\bibinfo  {journal} {Science}\ }\textbf {\bibinfo {volume}
  {349}},\ \bibinfo {pages} {283--286} (\bibinfo {year} {2015})}\BibitemShut {NoStop}%
\bibitem [{\citenamefont {Tonomura}\ \emph {et~al.}(2012)\citenamefont {Tonomura}, \citenamefont {Yu}, \citenamefont {Yanagisawa}, \citenamefont {Matsuda}, \citenamefont {Onose}, \citenamefont {Kanazawa}, \citenamefont {Park},\ and\ \citenamefont {Tokura}}]{tonomura_real-space_2012}%
  \BibitemOpen
  \bibfield  {author} {\bibinfo {author} {\bibfnamefont {A.}~\bibnamefont {Tonomura}}, \bibinfo {author} {\bibfnamefont {X.}~\bibnamefont {Yu}}, \bibinfo {author} {\bibfnamefont {K.}~\bibnamefont {Yanagisawa}}, \bibinfo {author} {\bibfnamefont {T.}~\bibnamefont {Matsuda}}, \bibinfo {author} {\bibfnamefont {Y.}~\bibnamefont {Onose}}, \bibinfo {author} {\bibfnamefont {N.}~\bibnamefont {Kanazawa}}, \bibinfo {author} {\bibfnamefont {H.~S.}\ \bibnamefont {Park}}, \ and\ \bibinfo {author} {\bibfnamefont {Y.}~\bibnamefont {Tokura}},\ }\bibfield  {title} {\enquote {\bibinfo {title} {Real-space observation of skyrmion lattice in helimagnet {MnSi} thin samples},}\ }\href {\doibase 10.1021/nl300073m} {\bibfield  {journal} {\bibinfo  {journal} {Nano Lett.}\ }\textbf {\bibinfo {volume} {12}},\ \bibinfo {pages} {1673--1677} (\bibinfo {year} {2012})}\BibitemShut {NoStop}%
\bibitem [{\citenamefont {Everschor-Sitte}\ and\ \citenamefont {Sitte}(2014)}]{everschor-sitte_real-space_2014}%
  \BibitemOpen
  \bibfield  {author} {\bibinfo {author} {\bibfnamefont {K.}~\bibnamefont {Everschor-Sitte}}\ and\ \bibinfo {author} {\bibfnamefont {M.}~\bibnamefont {Sitte}},\ }\bibfield  {title} {\enquote {\bibinfo {title} {Real-space {B}erry phases: Skyrmion soccer (invited)},}\ }\href {\doibase 10.1063/1.4870695} {\bibfield  {journal} {\bibinfo  {journal} {J. Appl. Phys.}\ }\textbf {\bibinfo {volume} {115}},\ \bibinfo {pages} {172602} (\bibinfo {year} {2014})}\BibitemShut {NoStop}%
\bibitem [{\citenamefont {Jiang}\ \emph {et~al.}(2017)\citenamefont {Jiang}, \citenamefont {Zhang}, \citenamefont {Yu}, \citenamefont {Zhang}, \citenamefont {Wang}, \citenamefont {Jungfleisch}, \citenamefont {Pearson}, \citenamefont {Cheng}, \citenamefont {Heinonen}, \citenamefont {Wang}, \citenamefont {Zhou}, \citenamefont {Hoffmann},\ and\ \citenamefont {te~Velthuis}}]{jiang_direct_2017}%
  \BibitemOpen
  \bibfield  {author} {\bibinfo {author} {\bibfnamefont {W.}~\bibnamefont {Jiang}}, \bibinfo {author} {\bibfnamefont {X.}~\bibnamefont {Zhang}}, \bibinfo {author} {\bibfnamefont {G.}~\bibnamefont {Yu}}, \bibinfo {author} {\bibfnamefont {W.}~\bibnamefont {Zhang}}, \bibinfo {author} {\bibfnamefont {X.}~\bibnamefont {Wang}}, \bibinfo {author} {\bibfnamefont {M.~B.}\ \bibnamefont {Jungfleisch}}, \bibinfo {author} {\bibfnamefont {J.~E.}\ \bibnamefont {Pearson}}, \bibinfo {author} {\bibfnamefont {X.}~\bibnamefont {Cheng}}, \bibinfo {author} {\bibfnamefont {O.}~\bibnamefont {Heinonen}}, \bibinfo {author} {\bibfnamefont {K.~L.}\ \bibnamefont {Wang}}, \bibinfo {author} {\bibfnamefont {Y.}~\bibnamefont {Zhou}}, \bibinfo {author} {\bibfnamefont {A.}~\bibnamefont {Hoffmann}}, \ and\ \bibinfo {author} {\bibfnamefont {S.~G.~E.}\ \bibnamefont {te~Velthuis}},\ }\bibfield  {title} {\enquote {\bibinfo {title} {Direct observation of the skyrmion {H}all effect},}\ }\href {\doibase 10.1038/NPHYS3883} {\bibfield  {journal}
  {\bibinfo  {journal} {Nature Phys.}\ }\textbf {\bibinfo {volume} {13}},\ \bibinfo {pages} {162--169} (\bibinfo {year} {2017})}\BibitemShut {NoStop}%
\bibitem [{\citenamefont {Iwasaki}\ \emph {et~al.}(2013{\natexlab{a}})\citenamefont {Iwasaki}, \citenamefont {Mochizuki},\ and\ \citenamefont {Nagaosa}}]{iwasaki_universal_2013}%
  \BibitemOpen
  \bibfield  {author} {\bibinfo {author} {\bibfnamefont {J.}~\bibnamefont {Iwasaki}}, \bibinfo {author} {\bibfnamefont {M.}~\bibnamefont {Mochizuki}}, \ and\ \bibinfo {author} {\bibfnamefont {N.}~\bibnamefont {Nagaosa}},\ }\bibfield  {title} {\enquote {\bibinfo {title} {Universal current-velocity relation of skyrmion motion in chiral magnets},}\ }\href {\doibase 10.1038/ncomms2442} {\bibfield  {journal} {\bibinfo  {journal} {Nature Commun.}\ }\textbf {\bibinfo {volume} {4}},\ \bibinfo {pages} {1463} (\bibinfo {year} {2013}{\natexlab{a}})}\BibitemShut {NoStop}%
\bibitem [{\citenamefont {Litzius}\ \emph {et~al.}(2017)\citenamefont {Litzius}, \citenamefont {Lemesh}, \citenamefont {Kr{\" u}ger}, \citenamefont {Bassirian}, \citenamefont {Caretta}, \citenamefont {Richter}, \citenamefont {B{\" u}ttner}, \citenamefont {Sato}, \citenamefont {Tretiakov}, \citenamefont {F{\" o}rster}, \citenamefont {Reeve}, \citenamefont {Weigand}, \citenamefont {Bykova}, \citenamefont {Stoll}, \citenamefont {Sch{\" u}tz}, \citenamefont {Beach},\ and\ \citenamefont {Kl{\" a}ui}}]{litzius_skyrmion_2017}%
  \BibitemOpen
  \bibfield  {author} {\bibinfo {author} {\bibfnamefont {K.}~\bibnamefont {Litzius}}, \bibinfo {author} {\bibfnamefont {I.}~\bibnamefont {Lemesh}}, \bibinfo {author} {\bibfnamefont {B.}~\bibnamefont {Kr{\" u}ger}}, \bibinfo {author} {\bibfnamefont {P.}~\bibnamefont {Bassirian}}, \bibinfo {author} {\bibfnamefont {L.}~\bibnamefont {Caretta}}, \bibinfo {author} {\bibfnamefont {K.}~\bibnamefont {Richter}}, \bibinfo {author} {\bibfnamefont {F.}~\bibnamefont {B{\" u}ttner}}, \bibinfo {author} {\bibfnamefont {K.}~\bibnamefont {Sato}}, \bibinfo {author} {\bibfnamefont {O.~A.}\ \bibnamefont {Tretiakov}}, \bibinfo {author} {\bibfnamefont {J.}~\bibnamefont {F{\" o}rster}}, \bibinfo {author} {\bibfnamefont {R.~M.}\ \bibnamefont {Reeve}}, \bibinfo {author} {\bibfnamefont {M.}~\bibnamefont {Weigand}}, \bibinfo {author} {\bibfnamefont {I.}~\bibnamefont {Bykova}}, \bibinfo {author} {\bibfnamefont {H.}~\bibnamefont {Stoll}}, \bibinfo {author} {\bibfnamefont {G.}~\bibnamefont {Sch{\" u}tz}}, \bibinfo {author} {\bibfnamefont
  {G.~S.~D.}\ \bibnamefont {Beach}}, \ and\ \bibinfo {author} {\bibfnamefont {M.}~\bibnamefont {Kl{\" a}ui}},\ }\bibfield  {title} {\enquote {\bibinfo {title} {Skyrmion {H}all effect revealed by direct time-resolved {X}-ray microscopy},}\ }\href {\doibase 10.1038/NPHYS4000} {\bibfield  {journal} {\bibinfo  {journal} {Nature Phys.}\ }\textbf {\bibinfo {volume} {13}},\ \bibinfo {pages} {170--175} (\bibinfo {year} {2017})}\BibitemShut {NoStop}%
\bibitem [{\citenamefont {G{\" o}bel}\ \emph {et~al.}(2021)\citenamefont {G{\" o}bel}, \citenamefont {Mertig},\ and\ \citenamefont {Tretiakov}}]{gobel_beyond_2021}%
  \BibitemOpen
  \bibfield  {author} {\bibinfo {author} {\bibfnamefont {B.}~\bibnamefont {G{\" o}bel}}, \bibinfo {author} {\bibfnamefont {I.}~\bibnamefont {Mertig}}, \ and\ \bibinfo {author} {\bibfnamefont {O.~A.}\ \bibnamefont {Tretiakov}},\ }\bibfield  {title} {\enquote {\bibinfo {title} {Beyond skyrmions: Review and perspectives of alternative magnetic quasiparticles},}\ }\href {\doibase 10.1016/j.physrep.2020.10.001} {\bibfield  {journal} {\bibinfo  {journal} {Phys. Rep.}\ }\textbf {\bibinfo {volume} {895}},\ \bibinfo {pages} {1} (\bibinfo {year} {2021})}\BibitemShut {NoStop}%
\bibitem [{\citenamefont {Bogdanov}\ and\ \citenamefont {Hubert}(1999)}]{bogdanov_stability_1999}%
  \BibitemOpen
  \bibfield  {author} {\bibinfo {author} {\bibfnamefont {A.}~\bibnamefont {Bogdanov}}\ and\ \bibinfo {author} {\bibfnamefont {A.}~\bibnamefont {Hubert}},\ }\bibfield  {title} {\enquote {\bibinfo {title} {The stability of vortex-like structures in uniaxial ferromagnets},}\ }\href {\doibase 10.1016/S0304-8853(98)01038-5} {\bibfield  {journal} {\bibinfo  {journal} {J. Mag. Mag. Mater.}\ }\textbf {\bibinfo {volume} {195}},\ \bibinfo {pages} {182--192} (\bibinfo {year} {1999})}\BibitemShut {NoStop}%
\bibitem [{\citenamefont {Streubel}\ \emph {et~al.}(2015)\citenamefont {Streubel}, \citenamefont {Han}, \citenamefont {Im}, \citenamefont {Kronast}, \citenamefont {R{\" o}\ss{}ler}, \citenamefont {Radu}, \citenamefont {Abrudan}, \citenamefont {Lin}, \citenamefont {Schmidt}, \citenamefont {Fischer},\ and\ \citenamefont {Makarov}}]{streubel_manipulating_2015}%
  \BibitemOpen
  \bibfield  {author} {\bibinfo {author} {\bibfnamefont {R.}~\bibnamefont {Streubel}}, \bibinfo {author} {\bibfnamefont {L.}~\bibnamefont {Han}}, \bibinfo {author} {\bibfnamefont {M.-Y.}\ \bibnamefont {Im}}, \bibinfo {author} {\bibfnamefont {F.}~\bibnamefont {Kronast}}, \bibinfo {author} {\bibfnamefont {U.~K.}\ \bibnamefont {R{\" o}\ss{}ler}}, \bibinfo {author} {\bibfnamefont {F.}~\bibnamefont {Radu}}, \bibinfo {author} {\bibfnamefont {R.}~\bibnamefont {Abrudan}}, \bibinfo {author} {\bibfnamefont {G.}~\bibnamefont {Lin}}, \bibinfo {author} {\bibfnamefont {O.~G.}\ \bibnamefont {Schmidt}}, \bibinfo {author} {\bibfnamefont {P.}~\bibnamefont {Fischer}}, \ and\ \bibinfo {author} {\bibfnamefont {D.}~\bibnamefont {Makarov}},\ }\bibfield  {title} {\enquote {\bibinfo {title} {Manipulating topological states by imprinting non-collinear spin textures},}\ }\href {\doibase 10.1038/srep08787} {\bibfield  {journal} {\bibinfo  {journal} {Sci. Rep.}\ }\textbf {\bibinfo {volume} {5}},\ \bibinfo {pages} {8787} (\bibinfo {year}
  {2015})}\BibitemShut {NoStop}%
\bibitem [{\citenamefont {Hagemeister}\ \emph {et~al.}(2018)\citenamefont {Hagemeister}, \citenamefont {Siemens}, \citenamefont {R\'ozsa}, \citenamefont {Vedmedenko},\ and\ \citenamefont {Wiesendanger}}]{hagemeister_controlled_2018}%
  \BibitemOpen
  \bibfield  {author} {\bibinfo {author} {\bibfnamefont {J.}~\bibnamefont {Hagemeister}}, \bibinfo {author} {\bibfnamefont {A.}~\bibnamefont {Siemens}}, \bibinfo {author} {\bibfnamefont {L.}~\bibnamefont {R\'ozsa}}, \bibinfo {author} {\bibfnamefont {E.~Y.}\ \bibnamefont {Vedmedenko}}, \ and\ \bibinfo {author} {\bibfnamefont {R.}~\bibnamefont {Wiesendanger}},\ }\bibfield  {title} {\enquote {\bibinfo {title} {Controlled creation and stability of $k\ensuremath{\pi}$ skyrmions on a discrete lattice},}\ }\href {\doibase 10.1103/PhysRevB.97.174436} {\bibfield  {journal} {\bibinfo  {journal} {Phys. Rev. B}\ }\textbf {\bibinfo {volume} {97}},\ \bibinfo {pages} {174436} (\bibinfo {year} {2018})}\BibitemShut {NoStop}%
\bibitem [{\citenamefont {Zheng}\ \emph {et~al.}(2017)\citenamefont {Zheng}, \citenamefont {Li}, \citenamefont {Wang}, \citenamefont {Song}, \citenamefont {Jin}, \citenamefont {Wei}, \citenamefont {Kov\'acs}, \citenamefont {Zang}, \citenamefont {Tian}, \citenamefont {Zhang}, \citenamefont {Du},\ and\ \citenamefont {Dunin-Borkowski}}]{zheng_direct_2017}%
  \BibitemOpen
  \bibfield  {author} {\bibinfo {author} {\bibfnamefont {F.}~\bibnamefont {Zheng}}, \bibinfo {author} {\bibfnamefont {H.}~\bibnamefont {Li}}, \bibinfo {author} {\bibfnamefont {S.}~\bibnamefont {Wang}}, \bibinfo {author} {\bibfnamefont {D.}~\bibnamefont {Song}}, \bibinfo {author} {\bibfnamefont {C.}~\bibnamefont {Jin}}, \bibinfo {author} {\bibfnamefont {W.}~\bibnamefont {Wei}}, \bibinfo {author} {\bibfnamefont {A.}~\bibnamefont {Kov\'acs}}, \bibinfo {author} {\bibfnamefont {J.}~\bibnamefont {Zang}}, \bibinfo {author} {\bibfnamefont {M.}~\bibnamefont {Tian}}, \bibinfo {author} {\bibfnamefont {Y.}~\bibnamefont {Zhang}}, \bibinfo {author} {\bibfnamefont {H.}~\bibnamefont {Du}}, \ and\ \bibinfo {author} {\bibfnamefont {R.~E.}\ \bibnamefont {Dunin-Borkowski}},\ }\bibfield  {title} {\enquote {\bibinfo {title} {Direct imaging of a zero-field target skyrmion and its polarity switch in a chiral magnetic nanodisk},}\ }\href {\doibase 10.1103/PhysRevLett.119.197205} {\bibfield  {journal} {\bibinfo  {journal} {Phys. Rev.
  Lett.}\ }\textbf {\bibinfo {volume} {119}},\ \bibinfo {pages} {197205} (\bibinfo {year} {2017})}\BibitemShut {NoStop}%
\bibitem [{\citenamefont {Kolesnikov}\ \emph {et~al.}(2018)\citenamefont {Kolesnikov}, \citenamefont {Stebliy}, \citenamefont {Samardak},\ and\ \citenamefont {Ognev}}]{kolesnikov_skyrmionium_2018}%
  \BibitemOpen
  \bibfield  {author} {\bibinfo {author} {\bibfnamefont {A.~G.}\ \bibnamefont {Kolesnikov}}, \bibinfo {author} {\bibfnamefont {M.~E.}\ \bibnamefont {Stebliy}}, \bibinfo {author} {\bibfnamefont {A.~S.}\ \bibnamefont {Samardak}}, \ and\ \bibinfo {author} {\bibfnamefont {A.~V.}\ \bibnamefont {Ognev}},\ }\bibfield  {title} {\enquote {\bibinfo {title} {Skyrmionium - high velocity without the skyrmion {H}all effect},}\ }\href {\doibase 10.1038/s41598-018-34934-2} {\bibfield  {journal} {\bibinfo  {journal} {Sci. Rep.}\ }\textbf {\bibinfo {volume} {8}},\ \bibinfo {pages} {16966} (\bibinfo {year} {2018})}\BibitemShut {NoStop}%
\bibitem [{\citenamefont {Finazzi}\ \emph {et~al.}(2013)\citenamefont {Finazzi}, \citenamefont {Savoini}, \citenamefont {Khorsand}, \citenamefont {Tsukamoto}, \citenamefont {Itoh}, \citenamefont {Du\`o}, \citenamefont {Kirilyuk}, \citenamefont {Rasing},\ and\ \citenamefont {Ezawa}}]{finazzi_laser-induced_2013}%
  \BibitemOpen
  \bibfield  {author} {\bibinfo {author} {\bibfnamefont {M.}~\bibnamefont {Finazzi}}, \bibinfo {author} {\bibfnamefont {M.}~\bibnamefont {Savoini}}, \bibinfo {author} {\bibfnamefont {A.~R.}\ \bibnamefont {Khorsand}}, \bibinfo {author} {\bibfnamefont {A.}~\bibnamefont {Tsukamoto}}, \bibinfo {author} {\bibfnamefont {A.}~\bibnamefont {Itoh}}, \bibinfo {author} {\bibfnamefont {L.}~\bibnamefont {Du\`o}}, \bibinfo {author} {\bibfnamefont {A.}~\bibnamefont {Kirilyuk}}, \bibinfo {author} {\bibfnamefont {Th.}\ \bibnamefont {Rasing}}, \ and\ \bibinfo {author} {\bibfnamefont {M.}~\bibnamefont {Ezawa}},\ }\bibfield  {title} {\enquote {\bibinfo {title} {Laser-induced magnetic nanostructures with tunable topological properties},}\ }\href {\doibase 10.1103/PhysRevLett.110.177205} {\bibfield  {journal} {\bibinfo  {journal} {Phys. Rev. Lett.}\ }\textbf {\bibinfo {volume} {110}},\ \bibinfo {pages} {177205} (\bibinfo {year} {2013})}\BibitemShut {NoStop}%
\bibitem [{\citenamefont {Fujita}\ and\ \citenamefont {Sato}(2017)}]{fujita_ultrafast_2017}%
  \BibitemOpen
  \bibfield  {author} {\bibinfo {author} {\bibfnamefont {H.}~\bibnamefont {Fujita}}\ and\ \bibinfo {author} {\bibfnamefont {M.}~\bibnamefont {Sato}},\ }\bibfield  {title} {\enquote {\bibinfo {title} {Ultrafast generation of skyrmionic defects with vortex beams: Printing laser profiles on magnets},}\ }\href {\doibase 10.1103/PhysRevB.95.054421} {\bibfield  {journal} {\bibinfo  {journal} {Phys. Rev. B}\ }\textbf {\bibinfo {volume} {95}},\ \bibinfo {pages} {054421} (\bibinfo {year} {2017})}\BibitemShut {NoStop}%
\bibitem [{\citenamefont {Ishida}\ and\ \citenamefont {Kondo}(2020)}]{ishida_theoretical_2020}%
  \BibitemOpen
  \bibfield  {author} {\bibinfo {author} {\bibfnamefont {Y.}~\bibnamefont {Ishida}}\ and\ \bibinfo {author} {\bibfnamefont {K.}~\bibnamefont {Kondo}},\ }\bibfield  {title} {\enquote {\bibinfo {title} {Theoretical comparison between skyrmion and skyrmionium motions for spintronics applications},}\ }\href {\doibase 10.7567/1347-4065/ab5b6b} {\bibfield  {journal} {\bibinfo  {journal} {Japan. J. Appl. Phys.}\ }\textbf {\bibinfo {volume} {59}},\ \bibinfo {pages} {SGGI04} (\bibinfo {year} {2020})}\BibitemShut {NoStop}%
\bibitem [{\citenamefont {Souza}\ \emph {et~al.}(2025)\citenamefont {Souza}, \citenamefont {Vizarim}, \citenamefont {Reichhardt}, \citenamefont {Reichhardt},\ and\ \citenamefont {Venegas}}]{souza_skyrmionium_2025}%
  \BibitemOpen
  \bibfield  {author} {\bibinfo {author} {\bibfnamefont {J.~C.~Bellizotti}\ \bibnamefont {Souza}}, \bibinfo {author} {\bibfnamefont {N.~P.}\ \bibnamefont {Vizarim}}, \bibinfo {author} {\bibfnamefont {C.~J.~O.}\ \bibnamefont {Reichhardt}}, \bibinfo {author} {\bibfnamefont {C.}~\bibnamefont {Reichhardt}}, \ and\ \bibinfo {author} {\bibfnamefont {P.~A.}\ \bibnamefont {Venegas}},\ }\href {\doibase 10.48550/arXiv.2501.06325} {\enquote {\bibinfo {title} {Skyrmionium {Dynamics} and {Stability} on {One} {Dimensional} {Anisotropy} {Patterns}},}\ } (\bibinfo {year} {2025})\BibitemShut {NoStop}%
\bibitem [{\citenamefont {Zhang}\ \emph {et~al.}(2016)\citenamefont {Zhang}, \citenamefont {Xia}, \citenamefont {Zhou}, \citenamefont {Wang}, \citenamefont {Liu}, \citenamefont {Zhao},\ and\ \citenamefont {Ezawa}}]{zhang_control_2016}%
  \BibitemOpen
  \bibfield  {author} {\bibinfo {author} {\bibfnamefont {X.}~\bibnamefont {Zhang}}, \bibinfo {author} {\bibfnamefont {J.}~\bibnamefont {Xia}}, \bibinfo {author} {\bibfnamefont {Y.}~\bibnamefont {Zhou}}, \bibinfo {author} {\bibfnamefont {D.}~\bibnamefont {Wang}}, \bibinfo {author} {\bibfnamefont {X.}~\bibnamefont {Liu}}, \bibinfo {author} {\bibfnamefont {W.}~\bibnamefont {Zhao}}, \ and\ \bibinfo {author} {\bibfnamefont {M.}~\bibnamefont {Ezawa}},\ }\bibfield  {title} {\enquote {\bibinfo {title} {Control and manipulation of a magnetic skyrmionium in nanostructures},}\ }\href {\doibase 10.1103/PhysRevB.94.094420} {\bibfield  {journal} {\bibinfo  {journal} {Phys. Rev. B}\ }\textbf {\bibinfo {volume} {94}},\ \bibinfo {pages} {094420} (\bibinfo {year} {2016})}\BibitemShut {NoStop}%
\bibitem [{\citenamefont {Komineas}\ and\ \citenamefont {Papanicolaou}(2015{\natexlab{a}})}]{komineas_skyrmion_2015_1}%
  \BibitemOpen
  \bibfield  {author} {\bibinfo {author} {\bibfnamefont {Stavros}\ \bibnamefont {Komineas}}\ and\ \bibinfo {author} {\bibfnamefont {Nikos}\ \bibnamefont {Papanicolaou}},\ }\bibfield  {title} {\enquote {\bibinfo {title} {Skyrmion dynamics in chiral ferromagnets under spin-transfer torque},}\ }\href {\doibase 10.1103/PhysRevB.92.174405} {\bibfield  {journal} {\bibinfo  {journal} {Physical Review B}\ }\textbf {\bibinfo {volume} {92}},\ \bibinfo {pages} {174405} (\bibinfo {year} {2015}{\natexlab{a}})}\BibitemShut {NoStop}%
\bibitem [{\citenamefont {Komineas}\ and\ \citenamefont {Papanicolaou}(2015{\natexlab{b}})}]{komineas_skyrmion_2015}%
  \BibitemOpen
  \bibfield  {author} {\bibinfo {author} {\bibfnamefont {S.}~\bibnamefont {Komineas}}\ and\ \bibinfo {author} {\bibfnamefont {N.}~\bibnamefont {Papanicolaou}},\ }\bibfield  {title} {\enquote {\bibinfo {title} {Skyrmion dynamics in chiral ferromagnets},}\ }\href {\doibase 10.1103/PhysRevB.92.064412} {\bibfield  {journal} {\bibinfo  {journal} {Phys. Rev. B}\ }\textbf {\bibinfo {volume} {92}},\ \bibinfo {pages} {064412} (\bibinfo {year} {2015}{\natexlab{b}})}\BibitemShut {NoStop}%
\bibitem [{\citenamefont {Li}\ \emph {et~al.}(2018)\citenamefont {Li}, \citenamefont {Xia}, \citenamefont {Zhang}, \citenamefont {Ezawa}, \citenamefont {Kang}, \citenamefont {Liu}, \citenamefont {Zhou},\ and\ \citenamefont {Zhao}}]{li_dynamics_2018}%
  \BibitemOpen
  \bibfield  {author} {\bibinfo {author} {\bibfnamefont {S.}~\bibnamefont {Li}}, \bibinfo {author} {\bibfnamefont {J.}~\bibnamefont {Xia}}, \bibinfo {author} {\bibfnamefont {X.}~\bibnamefont {Zhang}}, \bibinfo {author} {\bibfnamefont {M.}~\bibnamefont {Ezawa}}, \bibinfo {author} {\bibfnamefont {W.}~\bibnamefont {Kang}}, \bibinfo {author} {\bibfnamefont {X.}~\bibnamefont {Liu}}, \bibinfo {author} {\bibfnamefont {Y.}~\bibnamefont {Zhou}}, \ and\ \bibinfo {author} {\bibfnamefont {W.}~\bibnamefont {Zhao}},\ }\bibfield  {title} {\enquote {\bibinfo {title} {Dynamics of a magnetic skyrmionum driven by spin waves},}\ }\href {\doibase 10.1063/1.5026632} {\bibfield  {journal} {\bibinfo  {journal} {Appl. Phys. Lett.}\ }\textbf {\bibinfo {volume} {112}},\ \bibinfo {pages} {142404} (\bibinfo {year} {2018})}\BibitemShut {NoStop}%
\bibitem [{\citenamefont {Xia}\ \emph {et~al.}(2020)\citenamefont {Xia}, \citenamefont {Zhang}, \citenamefont {Ezawa}, \citenamefont {Tretiakov}, \citenamefont {Hou}, \citenamefont {Wang}, \citenamefont {Zhao}, \citenamefont {Liu}, \citenamefont {Diep},\ and\ \citenamefont {Zhou}}]{xia_current-driven_2020}%
  \BibitemOpen
  \bibfield  {author} {\bibinfo {author} {\bibfnamefont {J.}~\bibnamefont {Xia}}, \bibinfo {author} {\bibfnamefont {X.}~\bibnamefont {Zhang}}, \bibinfo {author} {\bibfnamefont {M.}~\bibnamefont {Ezawa}}, \bibinfo {author} {\bibfnamefont {O.~A.}\ \bibnamefont {Tretiakov}}, \bibinfo {author} {\bibfnamefont {Z.}~\bibnamefont {Hou}}, \bibinfo {author} {\bibfnamefont {W.}~\bibnamefont {Wang}}, \bibinfo {author} {\bibfnamefont {G.}~\bibnamefont {Zhao}}, \bibinfo {author} {\bibfnamefont {X.}~\bibnamefont {Liu}}, \bibinfo {author} {\bibfnamefont {H.~T.}\ \bibnamefont {Diep}}, \ and\ \bibinfo {author} {\bibfnamefont {Y.}~\bibnamefont {Zhou}},\ }\bibfield  {title} {\enquote {\bibinfo {title} {Current-driven skyrmionium in a frustrated magnetic system},}\ }\href {\doibase 10.1063/5.0012706} {\bibfield  {journal} {\bibinfo  {journal} {Appl. Phys. Lett.}\ }\textbf {\bibinfo {volume} {117}},\ \bibinfo {pages} {012403} (\bibinfo {year} {2020})}\BibitemShut {NoStop}%
\bibitem [{\citenamefont {Souza}\ \emph {et~al.}(2024)\citenamefont {Souza}, \citenamefont {Vizarim}, \citenamefont {Reichhardt}, \citenamefont {Reichhardt},\ and\ \citenamefont {Venegas}}]{souza_comparing_2024}%
  \BibitemOpen
  \bibfield  {author} {\bibinfo {author} {\bibfnamefont {J.~C.~Bellizotti}\ \bibnamefont {Souza}}, \bibinfo {author} {\bibfnamefont {N.~P.}\ \bibnamefont {Vizarim}}, \bibinfo {author} {\bibfnamefont {C.~J.~O.}\ \bibnamefont {Reichhardt}}, \bibinfo {author} {\bibfnamefont {C.}~\bibnamefont {Reichhardt}}, \ and\ \bibinfo {author} {\bibfnamefont {P.~A.}\ \bibnamefont {Venegas}},\ }\href {\doibase 10.48550/arXiv.2412.02001} {\enquote {\bibinfo {title} {Comparing {Dynamics}, {Pinning} and {Ratchet} {Effects} for {Skyrmionium}, {Skyrmions}, and {Antiskyrmions}},}\ } (\bibinfo {year} {2024})\BibitemShut {NoStop}%
\bibitem [{\citenamefont {Zheng}\ \emph {et~al.}(2022)\citenamefont {Zheng}, \citenamefont {Kiselev}, \citenamefont {Yang}, \citenamefont {Kuchkin}, \citenamefont {Rybakov}, \citenamefont {Bl{\" u}gel},\ and\ \citenamefont {Dunin-Borkowski}}]{zheng_skyrmionantiskyrmion_2022}%
  \BibitemOpen
  \bibfield  {author} {\bibinfo {author} {\bibfnamefont {F.}~\bibnamefont {Zheng}}, \bibinfo {author} {\bibfnamefont {N.~S.}\ \bibnamefont {Kiselev}}, \bibinfo {author} {\bibfnamefont {L.}~\bibnamefont {Yang}}, \bibinfo {author} {\bibfnamefont {V.~M.}\ \bibnamefont {Kuchkin}}, \bibinfo {author} {\bibfnamefont {F.~N.}\ \bibnamefont {Rybakov}}, \bibinfo {author} {\bibfnamefont {S.}~\bibnamefont {Bl{\" u}gel}}, \ and\ \bibinfo {author} {\bibfnamefont {R.~E.}\ \bibnamefont {Dunin-Borkowski}},\ }\bibfield  {title} {\enquote {\bibinfo {title} {Skyrmion-antiskyrmion pair creation and annihilation in a cubic chiral magnet},}\ }\href {\doibase 10.1038/s41567-022-01638-4} {\bibfield  {journal} {\bibinfo  {journal} {Nature Phys.}\ }\textbf {\bibinfo {volume} {18}},\ \bibinfo {pages} {863--868} (\bibinfo {year} {2022})}\BibitemShut {NoStop}%
\bibitem [{\citenamefont {Wang}\ \emph {et~al.}(2025)\citenamefont {Wang}, \citenamefont {Qiu},\ and\ \citenamefont {Shen}}]{wang_nonvolatile_2025}%
  \BibitemOpen
  \bibfield  {author} {\bibinfo {author} {\bibfnamefont {S.}~\bibnamefont {Wang}}, \bibinfo {author} {\bibfnamefont {L.}~\bibnamefont {Qiu}}, \ and\ \bibinfo {author} {\bibfnamefont {K.}~\bibnamefont {Shen}},\ }\bibfield  {title} {\enquote {\bibinfo {title} {Nonvolatile current-induced topological charge imbalance of magnetic textures},}\ }\href {\doibase 10.1038/s42005-025-01965-x} {\bibfield  {journal} {\bibinfo  {journal} {Commun. Phys.}\ }\textbf {\bibinfo {volume} {8}},\ \bibinfo {pages} {43} (\bibinfo {year} {2025})}\BibitemShut {NoStop}%
\bibitem [{\citenamefont {Zhang}\ \emph {et~al.}(2024)\citenamefont {Zhang}, \citenamefont {Tang}, \citenamefont {Wu}, \citenamefont {Shi}, \citenamefont {Xu}, \citenamefont {Wang}, \citenamefont {Tian},\ and\ \citenamefont {Du}}]{zhang_stable_2024}%
  \BibitemOpen
  \bibfield  {author} {\bibinfo {author} {\bibfnamefont {Y.}~\bibnamefont {Zhang}}, \bibinfo {author} {\bibfnamefont {J.}~\bibnamefont {Tang}}, \bibinfo {author} {\bibfnamefont {Y.}~\bibnamefont {Wu}}, \bibinfo {author} {\bibfnamefont {M.}~\bibnamefont {Shi}}, \bibinfo {author} {\bibfnamefont {X.}~\bibnamefont {Xu}}, \bibinfo {author} {\bibfnamefont {S.}~\bibnamefont {Wang}}, \bibinfo {author} {\bibfnamefont {M.}~\bibnamefont {Tian}}, \ and\ \bibinfo {author} {\bibfnamefont {H.}~\bibnamefont {Du}},\ }\bibfield  {title} {\enquote {\bibinfo {title} {Stable skyrmion bundles at room temperature and zero magnetic field in a chiral magnet},}\ }\href {\doibase 10.1038/s41467-024-47730-6} {\bibfield  {journal} {\bibinfo  {journal} {Nature Commun.}\ }\textbf {\bibinfo {volume} {15}},\ \bibinfo {pages} {3391} (\bibinfo {year} {2024})}\BibitemShut {NoStop}%
\bibitem [{\citenamefont {Zheng}\ \emph {et~al.}(2018)\citenamefont {Zheng}, \citenamefont {Rybakov}, \citenamefont {Borisov}, \citenamefont {Song}, \citenamefont {Wang}, \citenamefont {Li}, \citenamefont {Du}, \citenamefont {Kiselev}, \citenamefont {Caron}, \citenamefont {Kovacs}, \citenamefont {Tian}, \citenamefont {Zhang}, \citenamefont {Bl{\" u}gel},\ and\ \citenamefont {Dunin-Borkowski}}]{zheng_experimental_2018}%
  \BibitemOpen
  \bibfield  {author} {\bibinfo {author} {\bibfnamefont {F.}~\bibnamefont {Zheng}}, \bibinfo {author} {\bibfnamefont {F.~N.}\ \bibnamefont {Rybakov}}, \bibinfo {author} {\bibfnamefont {A.~B.}\ \bibnamefont {Borisov}}, \bibinfo {author} {\bibfnamefont {D.}~\bibnamefont {Song}}, \bibinfo {author} {\bibfnamefont {S.}~\bibnamefont {Wang}}, \bibinfo {author} {\bibfnamefont {Z.-A.}\ \bibnamefont {Li}}, \bibinfo {author} {\bibfnamefont {H.}~\bibnamefont {Du}}, \bibinfo {author} {\bibfnamefont {N.~S.}\ \bibnamefont {Kiselev}}, \bibinfo {author} {\bibfnamefont {J.}~\bibnamefont {Caron}}, \bibinfo {author} {\bibfnamefont {A.}~\bibnamefont {Kovacs}}, \bibinfo {author} {\bibfnamefont {M.}~\bibnamefont {Tian}}, \bibinfo {author} {\bibfnamefont {Y.}~\bibnamefont {Zhang}}, \bibinfo {author} {\bibfnamefont {S.}~\bibnamefont {Bl{\" u}gel}}, \ and\ \bibinfo {author} {\bibfnamefont {R.~E.}\ \bibnamefont {Dunin-Borkowski}},\ }\bibfield  {title} {\enquote {\bibinfo {title} {Experimental observation of chiral magnetic bobbers in
  {B}20-type {FeGe}},}\ }\href {\doibase 10.1038/s41565-018-0093-3} {\bibfield  {journal} {\bibinfo  {journal} {Nature Nanotechnol.}\ }\textbf {\bibinfo {volume} {13}},\ \bibinfo {pages} {451--455} (\bibinfo {year} {2018})}\BibitemShut {NoStop}%
\bibitem [{\citenamefont {Mandru}\ \emph {et~al.}(2020)\citenamefont {Mandru}, \citenamefont {Y{\i}ld{\i}r{\i}m}, \citenamefont {Tomasello}, \citenamefont {Heistracher}, \citenamefont {Penedo}, \citenamefont {Giordano}, \citenamefont {Suess}, \citenamefont {Finocchio},\ and\ \citenamefont {Hug}}]{mandru_coexistence_2020}%
  \BibitemOpen
  \bibfield  {author} {\bibinfo {author} {\bibfnamefont {A.-O.}\ \bibnamefont {Mandru}}, \bibinfo {author} {\bibfnamefont {O.}~\bibnamefont {Y{\i}ld{\i}r{\i}m}}, \bibinfo {author} {\bibfnamefont {R.}~\bibnamefont {Tomasello}}, \bibinfo {author} {\bibfnamefont {P.}~\bibnamefont {Heistracher}}, \bibinfo {author} {\bibfnamefont {M.}~\bibnamefont {Penedo}}, \bibinfo {author} {\bibfnamefont {A.}~\bibnamefont {Giordano}}, \bibinfo {author} {\bibfnamefont {D.}~\bibnamefont {Suess}}, \bibinfo {author} {\bibfnamefont {G.}~\bibnamefont {Finocchio}}, \ and\ \bibinfo {author} {\bibfnamefont {H.~J.}\ \bibnamefont {Hug}},\ }\bibfield  {title} {\enquote {\bibinfo {title} {Coexistence of distinct skyrmion phases observed in hybrid ferromagnetic/ferrimagnetic multilayers},}\ }\href {\doibase 10.1038/s41467-020-20025-2} {\bibfield  {journal} {\bibinfo  {journal} {Nature Commun.}\ }\textbf {\bibinfo {volume} {11}},\ \bibinfo {pages} {6365} (\bibinfo {year} {2020})}\BibitemShut {NoStop}%
\bibitem [{\citenamefont {Sohn}\ \emph {et~al.}(2019)\citenamefont {Sohn}, \citenamefont {Liu},\ and\ \citenamefont {Smalyukh}}]{sohn_schools_2019}%
  \BibitemOpen
  \bibfield  {author} {\bibinfo {author} {\bibfnamefont {H.~R.~O.}\ \bibnamefont {Sohn}}, \bibinfo {author} {\bibfnamefont {C.~D.}\ \bibnamefont {Liu}}, \ and\ \bibinfo {author} {\bibfnamefont {I.~I.}\ \bibnamefont {Smalyukh}},\ }\bibfield  {title} {\enquote {\bibinfo {title} {Schools of skyrmions with electrically tunable elastic interactions},}\ }\href {\doibase 10.1038/s41467-019-12723-3} {\bibfield  {journal} {\bibinfo  {journal} {Nature Commun.}\ }\textbf {\bibinfo {volume} {10}},\ \bibinfo {pages} {4744} (\bibinfo {year} {2019})}\BibitemShut {NoStop}%
\bibitem [{\citenamefont {Coelho}\ \emph {et~al.}(2023)\citenamefont {Coelho}, \citenamefont {Zhao}, \citenamefont {Tasinkevych}, \citenamefont {Smalyukh},\ and\ \citenamefont {Telo~da Gama}}]{coelho_sculpting_2023}%
  \BibitemOpen
  \bibfield  {author} {\bibinfo {author} {\bibfnamefont {R.~C.~V.}\ \bibnamefont {Coelho}}, \bibinfo {author} {\bibfnamefont {H.}~\bibnamefont {Zhao}}, \bibinfo {author} {\bibfnamefont {M.}~\bibnamefont {Tasinkevych}}, \bibinfo {author} {\bibfnamefont {I.~I.}\ \bibnamefont {Smalyukh}}, \ and\ \bibinfo {author} {\bibfnamefont {M.~M.}\ \bibnamefont {Telo~da Gama}},\ }\bibfield  {title} {\enquote {\bibinfo {title} {Sculpting liquid crystal skyrmions with external flows},}\ }\href {\doibase 10.1103/PhysRevResearch.5.033210} {\bibfield  {journal} {\bibinfo  {journal} {Phys. Rev. Res.}\ }\textbf {\bibinfo {volume} {5}},\ \bibinfo {pages} {033210} (\bibinfo {year} {2023})}\BibitemShut {NoStop}%
\bibitem [{\citenamefont {Amaral}\ \emph {et~al.}(2015)\citenamefont {Amaral}, \citenamefont {Zhao}, \citenamefont {Sedahmed}, \citenamefont {Campante}, \citenamefont {Smalyukh}, \citenamefont {Tasinkevych}, \citenamefont {Telo~da Gama},\ and\ \citenamefont {Coelho}}]{amaral_liquid_2025}%
  \BibitemOpen
  \bibfield  {author} {\bibinfo {author} {\bibfnamefont {G.~N.~C.}\ \bibnamefont {Amaral}}, \bibinfo {author} {\bibfnamefont {H.}~\bibnamefont {Zhao}}, \bibinfo {author} {\bibfnamefont {M.}~\bibnamefont {Sedahmed}}, \bibinfo {author} {\bibfnamefont {T.}~\bibnamefont {Campante}}, \bibinfo {author} {\bibfnamefont {I.~I.}\ \bibnamefont {Smalyukh}}, \bibinfo {author} {\bibfnamefont {M.}~\bibnamefont {Tasinkevych}}, \bibinfo {author} {\bibfnamefont {M.~M.}\ \bibnamefont {Telo~da Gama}}, \ and\ \bibinfo {author} {\bibfnamefont {R.~C.~V.}\ \bibnamefont {Coelho}},\ }\bibfield  {title} {\enquote {\bibinfo {title} {Liquid crystal torons in {P}oiseuille-like flows},}\ }\href {\doibase 10.1038/s41598-024-83294-7} {\bibfield  {journal} {\bibinfo  {journal} {Sci. Rep.}\ }\textbf {\bibinfo {volume} {15}},\ \bibinfo {pages} {2684} (\bibinfo {year} {2015})}\BibitemShut {NoStop}%
\bibitem [{\citenamefont {Evans}(2018)}]{evans_atomistic_2018}%
  \BibitemOpen
  \bibfield  {author} {\bibinfo {author} {\bibfnamefont {Richard F~L}\ \bibnamefont {Evans}},\ }\bibfield  {title} {\enquote {\bibinfo {title} {Atomistic {Spin} {Dynamics}},}\ }in\ \href {\doibase 10.1007/978-3-319-50257-1_147-1} {\emph {\bibinfo {booktitle} {Handbook of {Materials} {Modeling}: {Applications}: {Current} and {Emerging} {Materials}}}},\ \bibinfo {editor} {edited by\ \bibinfo {editor} {\bibfnamefont {Wanda}\ \bibnamefont {Andreoni}}\ and\ \bibinfo {editor} {\bibfnamefont {Sidney}\ \bibnamefont {Yip}}}\ (\bibinfo  {publisher} {Springer International Publishing},\ \bibinfo {year} {2018})\ pp.\ \bibinfo {pages} {1--23}\BibitemShut {NoStop}%
\bibitem [{\citenamefont {Iwasaki}\ \emph {et~al.}(2013{\natexlab{b}})\citenamefont {Iwasaki}, \citenamefont {Mochizuki},\ and\ \citenamefont {Nagaosa}}]{iwasaki_current-induced_2013}%
  \BibitemOpen
  \bibfield  {author} {\bibinfo {author} {\bibfnamefont {Junichi}\ \bibnamefont {Iwasaki}}, \bibinfo {author} {\bibfnamefont {Masahito}\ \bibnamefont {Mochizuki}}, \ and\ \bibinfo {author} {\bibfnamefont {Naoto}\ \bibnamefont {Nagaosa}},\ }\bibfield  {title} {\enquote {\bibinfo {title} {Current-induced skyrmion dynamics in constricted geometries},}\ }\href {\doibase 10.1038/nnano.2013.176} {\bibfield  {journal} {\bibinfo  {journal} {Nature Nanotechnology}\ }\textbf {\bibinfo {volume} {8}},\ \bibinfo {pages} {742--747} (\bibinfo {year} {2013}{\natexlab{b}})}\BibitemShut {NoStop}%
\bibitem [{\citenamefont {Paul}\ \emph {et~al.}(2020)\citenamefont {Paul}, \citenamefont {Haldar}, \citenamefont {von Malottki},\ and\ \citenamefont {Heinze}}]{paul_role_2020}%
  \BibitemOpen
  \bibfield  {author} {\bibinfo {author} {\bibfnamefont {S.}~\bibnamefont {Paul}}, \bibinfo {author} {\bibfnamefont {S.}~\bibnamefont {Haldar}}, \bibinfo {author} {\bibfnamefont {S.}~\bibnamefont {von Malottki}}, \ and\ \bibinfo {author} {\bibfnamefont {S.}~\bibnamefont {Heinze}},\ }\bibfield  {title} {\enquote {\bibinfo {title} {Role of higher-order exchange interactions for skyrmion stability},}\ }\href {\doibase 10.1038/s41467-020-18473-x} {\bibfield  {journal} {\bibinfo  {journal} {Nature Commun.}\ }\textbf {\bibinfo {volume} {11}},\ \bibinfo {pages} {4756} (\bibinfo {year} {2020})}\BibitemShut {NoStop}%
\bibitem [{\citenamefont {Seki}\ and\ \citenamefont {Mochizuki}(2016)}]{seki_skyrmions_2016}%
  \BibitemOpen
  \bibfield  {author} {\bibinfo {author} {\bibfnamefont {Shinichiro}\ \bibnamefont {Seki}}\ and\ \bibinfo {author} {\bibfnamefont {Masahito}\ \bibnamefont {Mochizuki}},\ }\href {\doibase 10.1007/978-3-319-24651-2} {\emph {\bibinfo {title} {Skyrmions in {Magnetic} {Materials}}}}\ (\bibinfo  {publisher} {Springer International Publishing},\ \bibinfo {year} {2016})\ \bibinfo {note} {series Title: SpringerBriefs in Physics}\BibitemShut {NoStop}%
\bibitem [{\citenamefont {Gilbert}(2004)}]{gilbert_phenomenological_2004}%
  \BibitemOpen
  \bibfield  {author} {\bibinfo {author} {\bibfnamefont {T.~L.}\ \bibnamefont {Gilbert}},\ }\bibfield  {title} {\enquote {\bibinfo {title} {A phenomenological theory of damping in ferromagnetic materials},}\ }\href {\doibase 10.1109/TMAG.2004.836740} {\bibfield  {journal} {\bibinfo  {journal} {IEEE Trans. Mag.}\ }\textbf {\bibinfo {volume} {40}},\ \bibinfo {pages} {3443--3449} (\bibinfo {year} {2004})}\BibitemShut {NoStop}%
\bibitem [{\citenamefont {Boulle}\ \emph {et~al.}(2016)\citenamefont {Boulle}, \citenamefont {Vogel}, \citenamefont {Yang}, \citenamefont {Pizzini}, \citenamefont {Chaves}, \citenamefont {Locatelli}, \citenamefont {Mente{\c s}}, \citenamefont {Sala}, \citenamefont {Buda-Prejbeanu}, \citenamefont {Klein}, \citenamefont {Belmeguenai}, \citenamefont {Roussign{\' e}}, \citenamefont {Stashkevich}, \citenamefont {Ch{\' e}rif}, \citenamefont {Aballe}, \citenamefont {Foerster}, \citenamefont {Chshiev}, \citenamefont {Auffret}, \citenamefont {Miron},\ and\ \citenamefont {Gaudin}}]{boulle_room-temperature_2016}%
  \BibitemOpen
  \bibfield  {author} {\bibinfo {author} {\bibfnamefont {O.}~\bibnamefont {Boulle}}, \bibinfo {author} {\bibfnamefont {J.}~\bibnamefont {Vogel}}, \bibinfo {author} {\bibfnamefont {H.}~\bibnamefont {Yang}}, \bibinfo {author} {\bibfnamefont {S.}~\bibnamefont {Pizzini}}, \bibinfo {author} {\bibfnamefont {D.~de~Souza}\ \bibnamefont {Chaves}}, \bibinfo {author} {\bibfnamefont {A.}~\bibnamefont {Locatelli}}, \bibinfo {author} {\bibfnamefont {T.~O.}\ \bibnamefont {Mente{\c s}}}, \bibinfo {author} {\bibfnamefont {A.}~\bibnamefont {Sala}}, \bibinfo {author} {\bibfnamefont {L.~D.}\ \bibnamefont {Buda-Prejbeanu}}, \bibinfo {author} {\bibfnamefont {O.}~\bibnamefont {Klein}}, \bibinfo {author} {\bibfnamefont {M.}~\bibnamefont {Belmeguenai}}, \bibinfo {author} {\bibfnamefont {Y.}~\bibnamefont {Roussign{\' e}}}, \bibinfo {author} {\bibfnamefont {A.}~\bibnamefont {Stashkevich}}, \bibinfo {author} {\bibfnamefont {S.~M.}\ \bibnamefont {Ch{\' e}rif}}, \bibinfo {author} {\bibfnamefont {L.}~\bibnamefont {Aballe}}, \bibinfo
  {author} {\bibfnamefont {M.}~\bibnamefont {Foerster}}, \bibinfo {author} {\bibfnamefont {M.}~\bibnamefont {Chshiev}}, \bibinfo {author} {\bibfnamefont {S.}~\bibnamefont {Auffret}}, \bibinfo {author} {\bibfnamefont {I.~M.}\ \bibnamefont {Miron}}, \ and\ \bibinfo {author} {\bibfnamefont {G.}~\bibnamefont {Gaudin}},\ }\bibfield  {title} {\enquote {\bibinfo {title} {Room-temperature chiral magnetic skyrmions in ultrathin magnetic nanostructures},}\ }\href {\doibase 10.1038/nnano.2015.315} {\bibfield  {journal} {\bibinfo  {journal} {Nature Nanotechnol.}\ }\textbf {\bibinfo {volume} {11}},\ \bibinfo {pages} {449--454} (\bibinfo {year} {2016})}\BibitemShut {NoStop}%
\bibitem [{\citenamefont {Kiefer}\ and\ \citenamefont {Wolfowitz}(1952)}]{kiefer_stochastic_1952}%
  \BibitemOpen
  \bibfield  {author} {\bibinfo {author} {\bibfnamefont {J.}~\bibnamefont {Kiefer}}\ and\ \bibinfo {author} {\bibfnamefont {J.}~\bibnamefont {Wolfowitz}},\ }\bibfield  {title} {\enquote {\bibinfo {title} {Stochastic estimation of the maximum of a regression function},}\ }\href {\doibase http://www.jstor.org/stable/2236690} {\bibfield  {journal} {\bibinfo  {journal} {Annal. Math. Stat.}\ }\textbf {\bibinfo {volume} {23}},\ \bibinfo {pages} {462--466} (\bibinfo {year} {1952})}\BibitemShut {NoStop}%
\bibitem [{\citenamefont {Robbins}\ and\ \citenamefont {Monro}(1951)}]{robbins_stochastic_1951}%
  \BibitemOpen
  \bibfield  {author} {\bibinfo {author} {\bibfnamefont {H.}~\bibnamefont {Robbins}}\ and\ \bibinfo {author} {\bibfnamefont {S.}~\bibnamefont {Monro}},\ }\bibfield  {title} {\enquote {\bibinfo {title} {A stochastic approximation method},}\ }\href {\doibase https://www.jstor.org/stable/2236626} {\bibfield  {journal} {\bibinfo  {journal} {Annal. Math. Stat.}\ }\textbf {\bibinfo {volume} {22}},\ \bibinfo {pages} {400--407} (\bibinfo {year} {1951})}\BibitemShut {NoStop}%
\bibitem [{\citenamefont {Lin}\ \emph {et~al.}(2013)\citenamefont {Lin}, \citenamefont {Reichhardt}, \citenamefont {Batista},\ and\ \citenamefont {Saxena}}]{lin_particle_2013}%
  \BibitemOpen
  \bibfield  {author} {\bibinfo {author} {\bibfnamefont {S.-Z.}\ \bibnamefont {Lin}}, \bibinfo {author} {\bibfnamefont {C.}~\bibnamefont {Reichhardt}}, \bibinfo {author} {\bibfnamefont {C.~D.}\ \bibnamefont {Batista}}, \ and\ \bibinfo {author} {\bibfnamefont {A.}~\bibnamefont {Saxena}},\ }\bibfield  {title} {\enquote {\bibinfo {title} {Particle model for skyrmions in metallic chiral magnets: Dynamics, pinning, and creep},}\ }\href {\doibase 10.1103/PhysRevB.87.214419} {\bibfield  {journal} {\bibinfo  {journal} {Phys. Rev. B}\ }\textbf {\bibinfo {volume} {87}},\ \bibinfo {pages} {214419} (\bibinfo {year} {2013})}\BibitemShut {NoStop}%
\bibitem [{\citenamefont {Raimondo}\ \emph {et~al.}(2022)\citenamefont {Raimondo}, \citenamefont {Saugar}, \citenamefont {Barker}, \citenamefont {Rodrigues}, \citenamefont {Giordano}, \citenamefont {Carpentieri}, \citenamefont {Jiang}, \citenamefont {Chubykalo-Fesenko}, \citenamefont {Tomasello},\ and\ \citenamefont {Finocchio}}]{raimondo_temperature-gradient-driven_2022}%
  \BibitemOpen
  \bibfield  {author} {\bibinfo {author} {\bibfnamefont {E.}~\bibnamefont {Raimondo}}, \bibinfo {author} {\bibfnamefont {E.}~\bibnamefont {Saugar}}, \bibinfo {author} {\bibfnamefont {J.}~\bibnamefont {Barker}}, \bibinfo {author} {\bibfnamefont {D.}~\bibnamefont {Rodrigues}}, \bibinfo {author} {\bibfnamefont {A.}~\bibnamefont {Giordano}}, \bibinfo {author} {\bibfnamefont {M.}~\bibnamefont {Carpentieri}}, \bibinfo {author} {\bibfnamefont {W.}~\bibnamefont {Jiang}}, \bibinfo {author} {\bibfnamefont {O.}~\bibnamefont {Chubykalo-Fesenko}}, \bibinfo {author} {\bibfnamefont {R.}~\bibnamefont {Tomasello}}, \ and\ \bibinfo {author} {\bibfnamefont {G.}~\bibnamefont {Finocchio}},\ }\bibfield  {title} {\enquote {\bibinfo {title} {Temperature-gradient-driven magnetic skyrmion motion},}\ }\href {\doibase 10.1103/PhysRevApplied.18.024062} {\bibfield  {journal} {\bibinfo  {journal} {Phys. Rev. Appl.}\ }\textbf {\bibinfo {volume} {18}},\ \bibinfo {pages} {024062} (\bibinfo {year} {2022})}\BibitemShut {NoStop}%
\bibitem [{\citenamefont {Zhang}\ \emph {et~al.}(2018)\citenamefont {Zhang}, \citenamefont {Wang}, \citenamefont {Burn}, \citenamefont {Peng}, \citenamefont {Berger}, \citenamefont {Bauer}, \citenamefont {Pfleiderer}, \citenamefont {van~der Laan},\ and\ \citenamefont {Hesjedal}}]{zhang_manipulation_2018}%
  \BibitemOpen
  \bibfield  {author} {\bibinfo {author} {\bibfnamefont {S.~L.}\ \bibnamefont {Zhang}}, \bibinfo {author} {\bibfnamefont {W.~W.}\ \bibnamefont {Wang}}, \bibinfo {author} {\bibfnamefont {D.~M.}\ \bibnamefont {Burn}}, \bibinfo {author} {\bibfnamefont {H.}~\bibnamefont {Peng}}, \bibinfo {author} {\bibfnamefont {H.}~\bibnamefont {Berger}}, \bibinfo {author} {\bibfnamefont {A.}~\bibnamefont {Bauer}}, \bibinfo {author} {\bibfnamefont {C.}~\bibnamefont {Pfleiderer}}, \bibinfo {author} {\bibfnamefont {G.}~\bibnamefont {van~der Laan}}, \ and\ \bibinfo {author} {\bibfnamefont {T.}~\bibnamefont {Hesjedal}},\ }\bibfield  {title} {\enquote {\bibinfo {title} {Manipulation of skyrmion motion by magnetic field gradients},}\ }\href {\doibase 10.1038/s41467-018-04563-4} {\bibfield  {journal} {\bibinfo  {journal} {Nature Commun.}\ }\textbf {\bibinfo {volume} {9}},\ \bibinfo {pages} {2115} (\bibinfo {year} {2018})}\BibitemShut {NoStop}%
\bibitem [{\citenamefont {Jiang}\ \emph {et~al.}(2024)\citenamefont {Jiang}, \citenamefont {Zhou}, \citenamefont {Zhang},\ and\ \citenamefont {Mochizuki}}]{jiang_transformation_2024}%
  \BibitemOpen
  \bibfield  {author} {\bibinfo {author} {\bibfnamefont {A.}~\bibnamefont {Jiang}}, \bibinfo {author} {\bibfnamefont {Y.}~\bibnamefont {Zhou}}, \bibinfo {author} {\bibfnamefont {X.}~\bibnamefont {Zhang}}, \ and\ \bibinfo {author} {\bibfnamefont {M.}~\bibnamefont {Mochizuki}},\ }\bibfield  {title} {\enquote {\bibinfo {title} {Transformation of a skyrmionium to a skyrmion through the thermal annihilation of the inner skyrmion},}\ }\href {\doibase 10.1103/PhysRevResearch.6.013229} {\bibfield  {journal} {\bibinfo  {journal} {Phys. Rev. Res.}\ }\textbf {\bibinfo {volume} {6}},\ \bibinfo {pages} {013229} (\bibinfo {year} {2024})}\BibitemShut {NoStop}%
\end{thebibliography}%
\end{document}